\pdfoutput=1
\documentclass[10pt]{article}

\usepackage[utf8]{inputenc}
\usepackage[T1]{fontenc}
\usepackage{textcomp}
\usepackage{authblk}
\usepackage{multirow}
\usepackage{array}
\usepackage{mathrsfs,amsmath}
\usepackage{float}
\usepackage{subfig}
\usepackage{color,soul}
\usepackage{subfloat}
\usepackage{lipsum,cuted}
\usepackage{verbatim}
\usepackage{graphicx}

\usepackage{tikz}
\usetikzlibrary{fit}
\usepackage[customcolors]{hf-tikz}
\usetikzlibrary{matrix,decorations.pathreplacing,calc}
\tikzset{style red/.style={
    set fill color=red!10,
    set border color=white,
  },
  style cyan/.style={
    set fill color=cyan!90!blue!60,
    set border color=white,
  },
  style orange/.style={
    set fill color=orange!80!red!60,
    set border color=white,
  },
  hor/.style={
    above left offset={-0.1,0.4},
    below right offset={0.1,-0.2},
    #1
  },
  ver/.style={
    above left offset={-0.1,0.3},
    below right offset={0.15,-0.15},
    #1
  }
}

\makeatletter
\newcommand*{\rom}[1]{\expandafter\@slowromancap\romannumeral #1@}
\makeatother

\title{High Efficiency Terahertz Generation in a Multi-Stage System} 

\author[1,2*]{Lu Wang}
\author[1]{Arya Fallahi}
\author[1,3]{Koustuban Ravi}
\author[1,2,4]{Franz K\"{a}rtner}
\affil[*]{lu.wang@desy.de}
\affil[1]{Center for Free Electron Laser Science, Deutsches Elektronen-Synchrotron Notkestra\text{$\beta$}e 85, 22607 Hamburg}
\affil[2]{Department of Physics, Universit\"{a}t of Hamburg, Luruper Chaussee 149, 22761 Hamburg, Germany}
\affil[3]{Research Laboratory of Electronics, Massachusetts Institute of Technology (MIT), Cambridge, Massachusetts 02139, USA}
\affil[4]{The Hamburg Centre for Ultrafast Imaging (CUI), Luruper Chaussee 149, 22761 Hamburg, Germany
}

\begin{document}
\maketitle

\newcommand{\degree}{^{\circ}}

\begin{abstract} 

We describe a robust system for laser-driven narrowband terahertz generation
with high conversion efficiency in periodically poled Lithium Niobate (PPLN). In the multi-stage terahertz generation system, the pump pulse is recycled after each PPLN stage for further terahertz generation. By out-coupling the terahertz radiation generated in each stage, extra absorption is circumvented and effective interaction length is increased. The separation of the terahertz and optical pulses at each stage is accomplished by an appropriately designed out-coupler. To evaluate the proposed architecture, the governing 2-D coupled wave equations in a cylindrically symmetric geometry are numerically solved using the finite difference method. Compared to the 1-D calculation which cannot capture the self-focusing and diffraction effects, our 2-D numerical method captures the effects of difference frequency generation, self-phase modulation,  self-focusing, beam diffraction, dispersion and terahertz absorption. We found that the terahertz generation efficiency can be greatly enhanced by compensating the dispersion of the pump pulse after each stage. With a two-stage system, we predict the generation of a $17.6$ mJ terahertz pulse with total conversion efficiency $\eta_{\text{total}}=1.6\%$ at $0.3$\,THz using a 1.1\,J pump laser with a two-line spectrum centered at 1 \textmu m. The generation efficiency of each stage is above $0.8\%$ with the out-coupling efficiencies above $93.0\%$.
\end{abstract}

\bibliographystyle{plain}

\bibliography{li.bib}

\section{Introduction} 
The last decades have seen a surge in research studies on generation and applications of terahertz radiation, which typically refers to electromagnetic waves in the spectral range from 0.1 to 3\,THz.
Among the wide range of applications, spectroscopy \cite{markelz2000pulsed,nagel2002integrated,young2017broadband}, spin dynamics control\cite{kampfrath2011coherent,kubacka2014large} and linear electron acceleration highly benefit from high power terahertz sources \cite{nanni2015terahertz,wong2013compact,zhang2018segmented}. In particular, applications like particle accelerations place steep requirements of a few millijoules of single/multi-cycle terahertz pulse energy to enable bunch manipulation in the relativistic regime. Such performance is largely contingent on increasing the terahertz generation efficiency towards the percent level and beyond. 

Different approaches have addressed the generation of few-cycle terahertz pulses, including four wave mixing in ambient air \cite{xie2006coherent}, plasma driven effects in air \cite{kress2004terahertz}, terahertz emission from photoconductive switches \cite{jepsen1996generation,dreyhaupt2005high}, tilted-pulse-fronts \cite{hebling2002velocity} and  echelons \cite{ofori2016thz,ravi2017broadband,palfalvi2017numerical}. In addition, there are techniques available for the generation of multi-cycle terahertz radiation.
Quantum cascade lasers are attractive due to their compactness and production simplicity in its use in the spectral range $>1$\,THz despite their limited tunability \cite{vijayraghavan2013broadly,williams2007terahertz,belkin2007terahertz}. Molecular gas lasers can provide high terahertz energies but are less tunable in terms of the generated terahertz frequency \cite{lee2009principles}. Free electron based terahertz sources are ideal in providing high energy multi-cycle terahertz pulses but are less accessible and are even more challenging for a compact implementation \cite{hauri2017intense,FLASH,wu2014thz}. Laser based multi-cycle terahertz generation can leverage on developments in solid-state laser technology to enable compact coherent terahertz sources with high conversion efficiency at low terahertz frequency (<1\,THz) \cite{vodopyanov2006optical,chen2011generation,tonouchi2007cutting}.  
This study revolves around the generation of high energy terahertz pulses in the multi-cycle (narrowband) regime.

Laser-driven terahertz generation utilizes the second order nonlinearity of the nonlinear material to perform difference frequency generation. GaAs, CdTe, InP \cite{vodopyanov2006terahertz,rice1994terahertz,vodopyanov2008optical}  HMQ-TMS \cite{lu2015tunable}, LiNbO\textsubscript{3} \cite{lee2000generation}, GaP, ZnTe \cite{dietze2012dynamically} and DAST \cite{kawase2000tunable,schneider2006generation} are nonlinear materials that have received significant attention. Multi-cycle terahertz generation in periodically poled Lithium Niobate (PPLN) was initially demonstrated by Lee et al. \cite{lee2000generation}. Recent research efforts focused on terahertz generation in PPLNs with various pump pulse formats such as chirp and delay, pulse-train and cascaded optical parametric amplification\cite{vodopyanov2006optical,ravi2016pulse,cirmi2017cascaded}. To our knowledge, the highest experimentally achieved conversion efficiency for multi-cycle terahertz generation in PPLN is $0.13\%$  \cite{ahr2017narrowband}.
 
Increasing the optical to terahertz generation/conversion efficiency (referred to as \emph{efficiency} throughout this paper) in laser-based schemes is often hindered by several challenges.
The first limitation is introduced by the Manley-Rowe relation \cite{manley1956some} obtained after the assumption of single photon conversion.
This limitation can be overcome by a cascaded second order process (repeated energy down-conversion of pump photons
to terahertz photons) \cite{cirmi2017cascaded,cronin2004cascaded}. The damage threshold of the nonlinear crystal imposes another challenge. Large pump electric field strength induces optical breakdown and damage to the material, which also limits the achievable terahertz efficiency \cite{hoffmann2007efficient}.
Another challenge is introduced by the terahertz absorption of the material. This challenge can be overcome by cryogenic cooling \cite{shikata1999enhancement,huang2013high}, since the temperature decrease impedes thermal phonon excitation and, therefore, considerably reduces the terahertz absorption. Additionally, it was found by Lee et al. that cooling of the nonlinear crystal increases the terahertz generation efficiency \cite{lee2000temperature}.

Our previous work based on a 1D-propagation model \cite{ravi2016pulse}, partially addressed these issues and showed the possibility of obtaining high efficiencies of several percent for optimal pump pulse formats. However, in practice, complex pump pulse formats are hard to engineer. Therefore, further innovations are necessary to achieve the desired percent level efficiencies. Here, we aim to alleviate these limitations by recycling the optical pump and separating the generated terahertz radiation in a staged approach, before significant terahertz absorption occurs. In this multi-stage architecture, consecutive stages utilize the same optical pump pulse. The terahertz pulse generated at each PPLN stage is then coupled out, circumventing the excessive terahertz absorption.
A similar concept was experimentally tested by Chen et al. \cite{chen2008generation} with a two-stage system for single-cycle terahertz generation. However, in their system, the terahertz pulse generated in the first stage is also coupled into the second stage together with the pump pulse \cite{chen2008generation}.

 In this paper, section 2 describes the coupled wave equations governing terahertz generation and the utilized numerical approach based on a finite difference method in cylindrical coordinates. The
effects of dispersion and mitigation mechanisms are discussed. Furthermore, the critical aspect of isolating the co-propagating terahertz and optical beams at each stage is resolved with a suitably designed quartz coupler. Section 3 analyzes the dependence of terahertz generation on pump pulse duration, effective length and beam size. Also, the influence of phase front distortions and spatial intensity fluctuations of the pump pulse on terahertz generation are studied. Detailed information  about the terahertz spatial profile and terahertz beam combination  is presented.

\section{Multi-stage Terahertz Generation: Concept and Physics}
\begin{figure}[H]
\centering{  
{ 
   \includegraphics[width=0.65\textwidth]{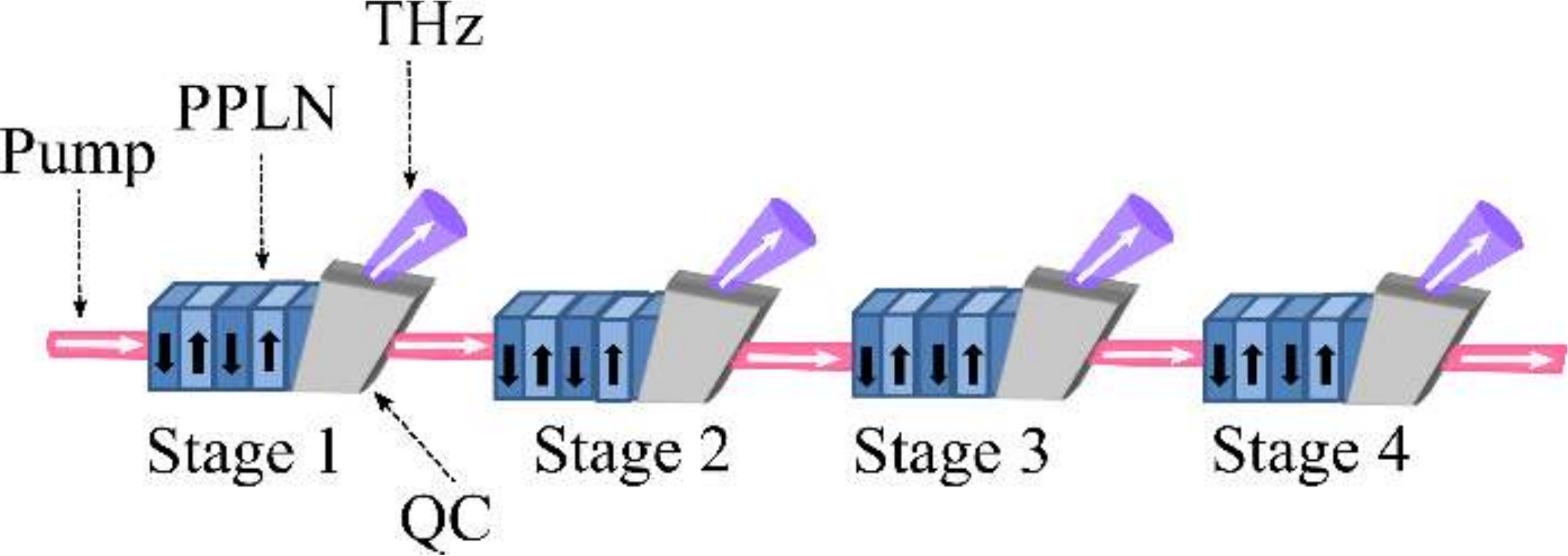}}

   \caption{Schematic illustration of the multi-stage system. Detailed information about the quartz coupler (QC) can be found in Fig.\,\ref{brewster_angel1}. In the sketch, the angular deflection of the optical pump caused by the refraction at the output of each stage is not delineated for illustration purposes. The dark arrows on the PPLN crystals represent the poling of the nonlinear material. }\label{setup_sketch_1}}
\end{figure}
The schematic illustration of the multi-stage system is presented in Fig.\,\ref{setup_sketch_1}. A set of PPLN crystals are placed in a serial arrangement, through which an optical pump pulse is consecutively recycled. At each stage, the terahertz beam is out-coupled by a designed quartz coupler (see section \ref{my_coupler}) to avoid excessive terahertz absorption. The out-coupling and separation of the optical and terahertz beams may also be achieved by a combination of anti-reflection coatings for optical and terahertz beams at the exit of the crystal, and a dichroic mirror. In this article, we focus on the quartz coupler due to its simplicity in manufacturing and the low loss for the terahertz beam.     

\subsection{Theory}\label{my_num_method}

Terahertz radiation is generated in PPLN through difference frequency generation (DFG) processes. The collinear propagation of terahertz and pump beams suggest considering the cylindrical symmetry in solving the governing equations for terahertz generation. For the quartz coupler, however, this cylindrical symmetry is violated. We solve for the propagation of terahertz beam in quartz using an angular spectrum method \cite{goodman2005introduction}. For this purpose, the terahertz beam profile is reconstructed in 3-D Cartesian coordinates from the 2-D cylindrical coordinates. It is assumed that the quartz coupler has no influence on the pump pulse (detail in section \ref{set_up_sketch}).
To precisely simulate the evolution of the terahertz and optical pulses, all dominant effects such as DFG, self-phase modulation (SPM), self-focusing, beam diffraction, dispersion and material absorption of the terahertz radiation, are included.
In our studies, stimulated Raman scattering is neglected because of its relatively small influence \cite{ravi2014limitations}, since the desired terahertz frequency range (< 1 THz) is far away from the phonon resonance frequency of Lithium Niobate (LiNbO\textsubscript{3}), occurring around 7.5\,THz\cite{schwarz1997asymmetric}.

\subsubsection{Coupled Wave Equations in Cylindrical Coordinate}

\begin{figure}[H]
\centering{  
{ 
   \includegraphics[width=0.48\textwidth]{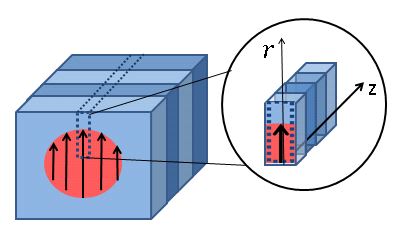}}

   \caption{Schematic illustration of the simulated geometry: the dark thick arrows represent the polarization direction of both pump and terahertz beams. The polarization direction is aligned with the extraordinary optical axis of PPLN.  The origin of the cylindrical coordinate is at the center of the beam. $r$ and $z$ represent the transverse and propagation directions, respectively.}\label{cylin_slice}}
\end{figure}

We choose to perform calculations on a slice along the radial axis in cylindrical coordinates as shown in Fig.\,\ref{cylin_slice}. 
For high energy terahertz generation, the usage of large crystals is mandatory due to the limited damage threshold of the nonlinear materials. Using the state-of-the-art fabrication technology, PPLN crystals as large as 1 cm $\times$ 1 cm transverse dimensions and few centimeters in length are realizable \cite{ishizuki2012half}. 
The propagation of the pump and terahertz beams in bulk PPLN crystals resemble the propagation of linearly polarized beams in a square waveguide.
However, the waveguide dimensions are far larger than the wavelength of both pump and terahertz in the crystal. As a result, the effect of the waveguide boundaries become negligible. 

Consequently, the entire problem reduces to solving the scalar coupled wave equations with the ansatz $E(\omega,r,z)=A(\omega,r,z)e^{-ik(\omega)z}$, considering the slowly varying amplitude approximation. 
The governing equations of the field amplitude of pump and terahertz beams read as:
\begin{eqnarray}
&&\frac{\partial A_{\text{op}}(\omega,r,z)}{\partial z}=\frac{-i}{2k(\omega)}\frac{1}{r}\frac{\partial}{\partial r}\left({r\frac{\partial A_{\mathrm{op}}(\omega,r,z)}{\partial r}}\right)-\frac{i\varepsilon_0 n(\omega)n_2 \omega}{2}\mathscr{F}\lbrace|A_\text{op}(t,r,z)|^2A_{\mathrm{op}}(t,r,z)\rbrace\nonumber \\
&&-\frac{i\omega \chi_{\text{eff}}(z)}{2n(\omega) c }\int_{-\infty}^\infty A_{\mathrm{op}}(\omega+\Omega,r,z)A^*_{\mathrm{THz}}(\Omega,r,z)e^{i\bigtriangleup kz}d\Omega \label{ir_cylin} \\
\
\ \nonumber \\
&&\frac{\partial A_{\mathrm{THz}}(\Omega,r,z)}{\partial z}=\frac{-i}{2k(\Omega)}\frac{1}{r}\frac{\partial}{\partial r}\left({r\frac{\partial A_{\mathrm{THz}}(\Omega,r,z)}{\partial r}}\right)-\frac{\alpha(\Omega)}{2}A_{\mathrm{THz}}(\Omega,r,z)\nonumber \\
&&-\frac{i\Omega \chi_{\text{eff}}(z)}{2n(\Omega) c }\int_{-\infty}^\infty A_\mathrm{op}(\omega+\Omega,r,z)A^*_\mathrm{op}(\omega,r,z)e^{i\bigtriangleup kz}d\omega \label{thz_cylin}
\end{eqnarray}

In Eqs. (\ref{ir_cylin}) and (\ref{thz_cylin}), $A_\mathrm{op}$ and $A_\mathrm{THz}$ stand for the electric field amplitude of the optical pump and the terahertz beam respectively; $\mathscr{F}$ denotes the Fourier transform operator; $\omega ~\text{and}~ \Omega$ are the optical and terahertz angular frequencies respectively; $k$ represents the wavenumber; $\alpha$ denotes terahertz absorption; $n_2$ represents the nonlinear refractive index and $n$ represents the refractive index. The effective nonlinear coefficient is written as $\chi^{(2)}_{\text{eff}}=\chi^{(2)}_{333}(z)$ in the last terms on the right hand side of Eqs.\,(1) and (2). The term $e^{i\bigtriangleup kz}=e^{i(k(\Omega)+k(\omega)-k(\omega+\Omega))z}$ evaluates phase mismatch between the propagation of the terahertz beam and the optical pump. The term $e^{i\bigtriangleup kz}$ together with the term in the first order Fourier expansion series of $\chi^{(2)}_{333}(z)$, leads to the phase-matching condition \cite{ravi2016pulse}.

 On the right hand side of Eqs.\,(\ref{ir_cylin}) and (2), the first terms correspond to the transverse spatial dependence of the propagation, describing beam divergence due to diffraction. In Eq.\,(\ref{ir_cylin}), the second term describes the impact of SPM on the optical beam which is independent of the phase-matching condition. As one can see, we consider a frequency dependent SPM effect which implicitly includes the effect of self-steepening (see Appendix \ref{f_spm}). Additionally, by considering radially dependent SPM, self-focusing is considered.  The third term captures two important effects. One is the DFG term $A_\mathrm{op}(\omega+\Omega)A^*_\mathrm{THz}(\Omega)$ term which describes the frequency down-shift of the pump, and the other is the $A_\mathrm{op}(\omega-\Omega)A_\mathrm{THz}(\Omega)$ term which describes the back conversion from terahertz to the pump beam (sum-frequency generation)\cite{ravi2014theory}. The second and third terms on the right hand side of Eq.\,(\ref{thz_cylin}) evaluate the effects of terahertz absorption and DFG processes respectively.

 Our 2-D calculations in cylindrical coordinates allow for the possibility of including the spatial variation of the beam, diffraction and self-focusing. This can not be addressed by a 1-D calculation.

\subsubsection{Numerical Method and Simulation Parameters}\label{numer_method}
Various finite difference methods can be used for solving Eqs.\,(\ref{ir_cylin}) and (\ref{thz_cylin}). In this study, a split step Fourier method is used to compute the integral terms corresponding to the nonlinear polarization terms containing $n_2$ and $\chi^{(2)}_{\text{eff}}$. The second order derivative in the $r$ direction is expressed by a conventional 3-point finite difference method as shown by Eq.\,(\ref{discretization_r}) in Appendix (\ref{numerical_method_ap}). Ultimately, the values of $A_\mathrm{op}$ and $A_\mathrm{THz}$ are updated throughout the propagation distance $z$ using the low-storage Runge-Kutta method  \cite{williamson1980low}. 
The entire method leads to an explicit update algorithm along $z$ rather than in time $t$. This is only valid when there is no reflection in the system (i.e only forward propagation along $z$ is involved).
The 3-point finite difference method takes advantage of the tridiagonal matrix of the discretization along the $r$ direction, thus reducing computational cost. 
A combination of split step, explicit finite difference, and low-storage Runge-Kutta methods highly enhance the computational performance. Using other finite difference methods such as the Crank-Nicolson method is also possible.
However, this implicit method necessitates the inversion of a finite difference matrix for each frequency and increases computational cost. Additionally, an angular spectrum method based on Hankel transformation \cite{siegman1977quasi} can be used but the approach suffers from a lack of the development of the fast Fourier transform (FFT) in Bessel series \cite{cree1993algorithms} and the boundary conditions are less straightforward to be implemented. Our calculation is performed in C++ with MPI and OpenMP for code parallelization. Details of the numerical method can be found in Appendix (\ref{numerical_method_ap}).

In this work, we base all simulations for terahertz generation centered at 0.3\,THz due to its amenability for electron acceleration. The overarching conclusions may however be extended to other frequencies as well. Terahertz generation with pulse trains is effective in mitigating walk-off and yields high conversion efficiencies\cite{ravi2016pulse}. In the remainder of this paper, the simulations are based on pump pulses consisting of two narrow spectral lines (referred to as \emph{two-lines} throughout this paper) separated by the desired terahertz frequency ($\Omega_0=2\pi\times$0.3\,THz). This forms a pulse train (beat note) with each sub-pulse separated by a time $2\pi/\Omega_0$, i.e. the terahertz electric field period. The simulation parameters are tabulated in Table (\ref{parameter_table}). 
\bgroup
\def\arraystretch{1.5}
\begin{table}[ht!]
\centering{\caption{Simulation parameters with two-lines input}\label{parameter_table}}
\begin{tabular}{|l| l|   }
\hline
\textbf{Parameters} & \textbf{Values} \\
\hline
$n_2$ \cite{desalvo1996infrared}  & $1.25\times10^{-19}$\,$\mathrm{W/m^2}$\\
\hline
 $\chi^{(2)}_{\text{eff}}=\chi^{(2)}_{333}$ \cite{weis1985lithium,vodopyanov2006optical,hebling2008generation,ravi2016pulse}& $2\times 168\,$pm/V\\
\hline
Super-Gaussian Order M & 5 \\
\hline
 $\Omega_{0}$ & $2\pi\times0.3\,\text{THz}$ \\
 \hline

PPLN Period $\Lambda$ &374.1\,\textmu m\\
\hline 
$\alpha(\Omega_0)$\cite{wu2015temperature}& $1.4\,/\text{cm}$\\
\hline
Damage Fluence $F_\text{d}$ \cite{ravi2016pulse} &
 $10\sqrt{\tau_{\text{FWHM}}/(2\times 10\,\text{ns})}$ $\mathrm{J/{cm}^2}$ \\
 \hline
  Temperature  & 75\,K \\
\hline

\multicolumn{2}{|c|}{
\textbf{Input Pump Pulse Electric Field Format}
}\\
\hline
\multicolumn{2}{|c|}{
Two-lines frequencies ($\omega_0,\omega_0+\Omega_0$)     291.26\,THz, 291.56\,THz }\\
\hline
\multicolumn{2}{|c|}{
 $E(\omega) \propto   e^{-(\omega-\omega_0)^2\tau_{\text{FWHM}}^2/(8\ln{2})}+e^{-(\omega-\omega_0-\Omega_0)^2\tau_{\text{FWHM}}^2/{(8\ln{2})}}$} \\
 \hline
 \multicolumn{2}{|c|}{
 $E(t) \propto e^{i\omega_0 t}e^{-2\ln{2}\big(t/{\tau_{\text{FWHM}}}\big)^2}[1+ e^{i\Omega_0 t}]$ } \\
 \hline
 \multicolumn{2}{|c|}{
$E(r) \propto e^{(-r^2/2 \sigma ^2)^M} $}\\
 \hline
\end{tabular}
\end{table}
\egroup

The refractive index and absorption ($\alpha$) of terahertz radiation at $80$\,K and 0.3\,THz are obtained from linear interpolation of the experimental data between temperatures $50$\,K and $100$\,K from \cite{wu2015temperature}.
The material absorption at optical wavelengths, i.e. the pump absorption, is neglected and the refractive index is fitted to a Sellmeier equation \cite{jundt1997temperature}.
We assume a super-Gaussian spatial distribution ($M=5$) and a Gaussian temporal profile for the optical pump. This is also the case for high power lasers. Additionally, a super-Gaussian spatial distribution ensures that the generated terahertz radiation is homogeneous in the transverse $r$ dimension and also minimizes the damage due to peak fluence of the pump compared to a Gaussian spatial profile. The input pump pulse energy, which follows from spatial and temporal integrations of the electric field defined in Table.\,1, is given by
\begin{equation}\label{energy_input}
\text{Energy}=\frac{2\pi \sigma ^2 \Gamma (\frac{M+1}{M}) F_\text{d}}{2^{\frac{1}{M}}} 
\end{equation}
where $\Gamma$ is the gamma function, $\sigma$ is the waist of the pump beam  and $\tau_{\text{FWHM}}$ is the full width half maximum (FWHM) of the pump temporal envelope. 

\subsection{Terahertz Efficiency Enhancement with Dispersion Compensation}

We suggest compensating the dispersion accumulated in the material of the pump pulse generated at the end of each stage before recycling it to the subsequent PPLN stage in order to enhance the conversion efficiency. In other words, before the pump pulse is recycled to stage N, an opposite dispersion is added to the pump pulse at the end of stage N-1 to cancel the dispersion obtained by propagation through the stage N-1.
For the first and second stages, dispersion compensation can be realized by a prism or grating (i.e second and third order dispersion). For many stages, the cascading process broadens the spectrum drastically. Dispersion compensation for a large bandwidth beam can be challenging in practice since higher order dispersion needs to be taken into consideration. However, it still can be achieved with double chirped mirrors \cite{kartner1997design}.

\begin{figure}[H]
  \centering

{\includegraphics[width=10.0cm]{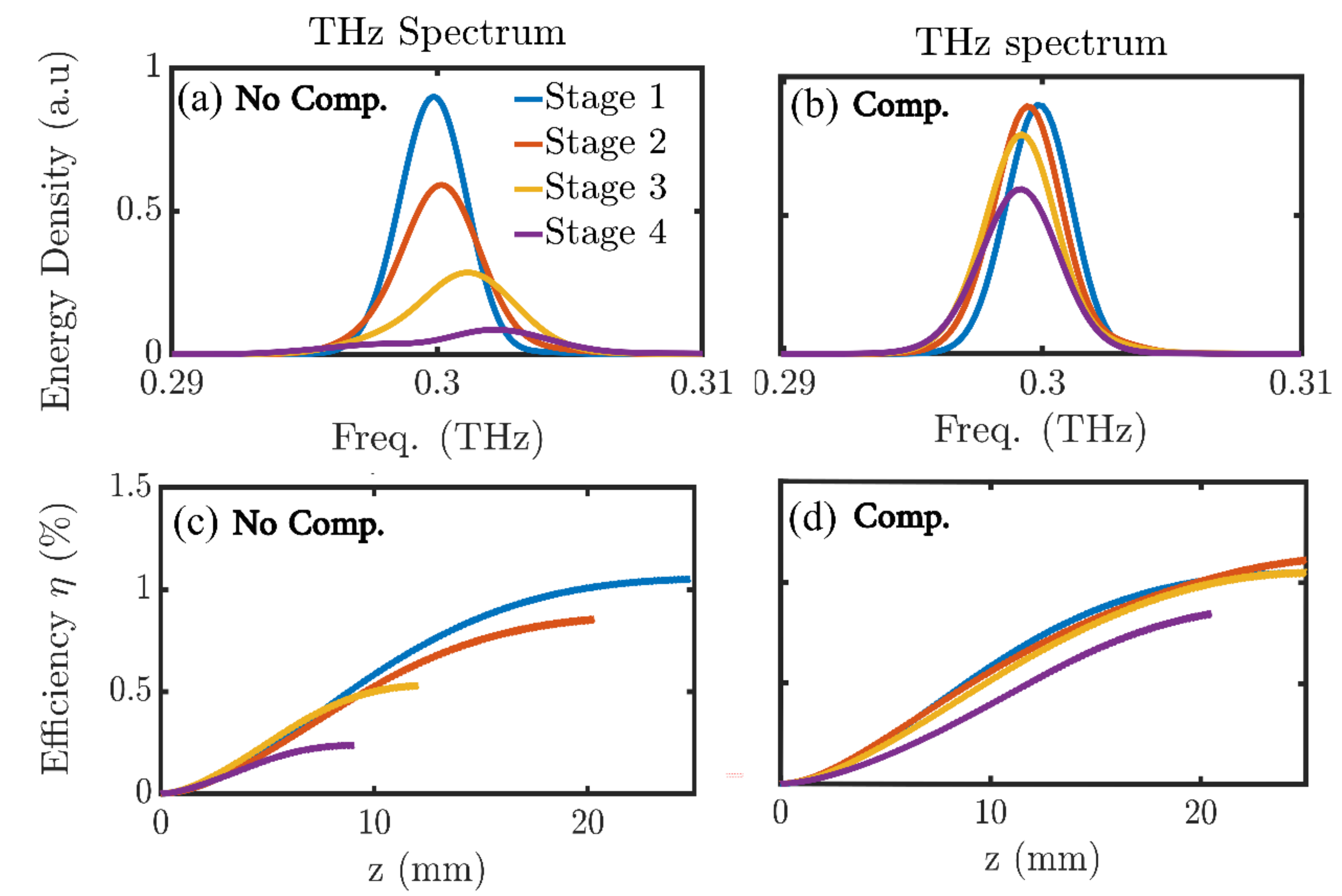}} 
 
\caption{Comparison of a 4 stage system with input pump pulse parameters $\sigma=5$\,mm, $\tau_{\text{FWHM}}=150$\,ps. The effective length of each stage is chosen to be the distance where efficiency saturates. (a,c) depict the terahertz spectra generated after each stage and the conversion efficiency by recycling the pump pulse directly without dispersion compensation (No Comp.). (b,d) show the terahertz spectra and the conversion efficiency with pump pulse dispersion compensation (Comp.) after each stage.\label{4_stages_eff1}}
\end{figure}

The simulation results comparing the terahertz spectra at the end of each stage with and without dispersion compensation are shown in Fig.\,\ref{4_stages_eff1}(a,b). The resulting conversion efficiencies are shown in Fig.\,\ref{4_stages_eff1}(c,d). 

One can see that the terahertz conversion efficiency reduces subsequently after each stage without dispersion compensation. Since
the terahertz generation broadens the pump spectrum via the cascading process, part of the broadened pump spectrum can not be phase matched for further cascading. In other words, different frequency elements in the pump spectrum pick up different phases due to dispersion in the PPLN via propagation. Consequently, the terahertz radiation generated by DFG from different spectral ranges of the pump pulse carries different phases, leading to partial destructive interference of the total generated terahertz radiation. Thus, the efficiency is degraded due to the dispersion. It can be seen in Fig.\,\ref{4_stages_eff1}(b,d) that the efficiency can be significantly enhanced by dispersion compensation. 

\subsubsection{Spectral Dynamics of Optical Pump and Terahertz Pulses}

\begin{figure}[H]
  \centering

{\includegraphics[width=12.5cm]{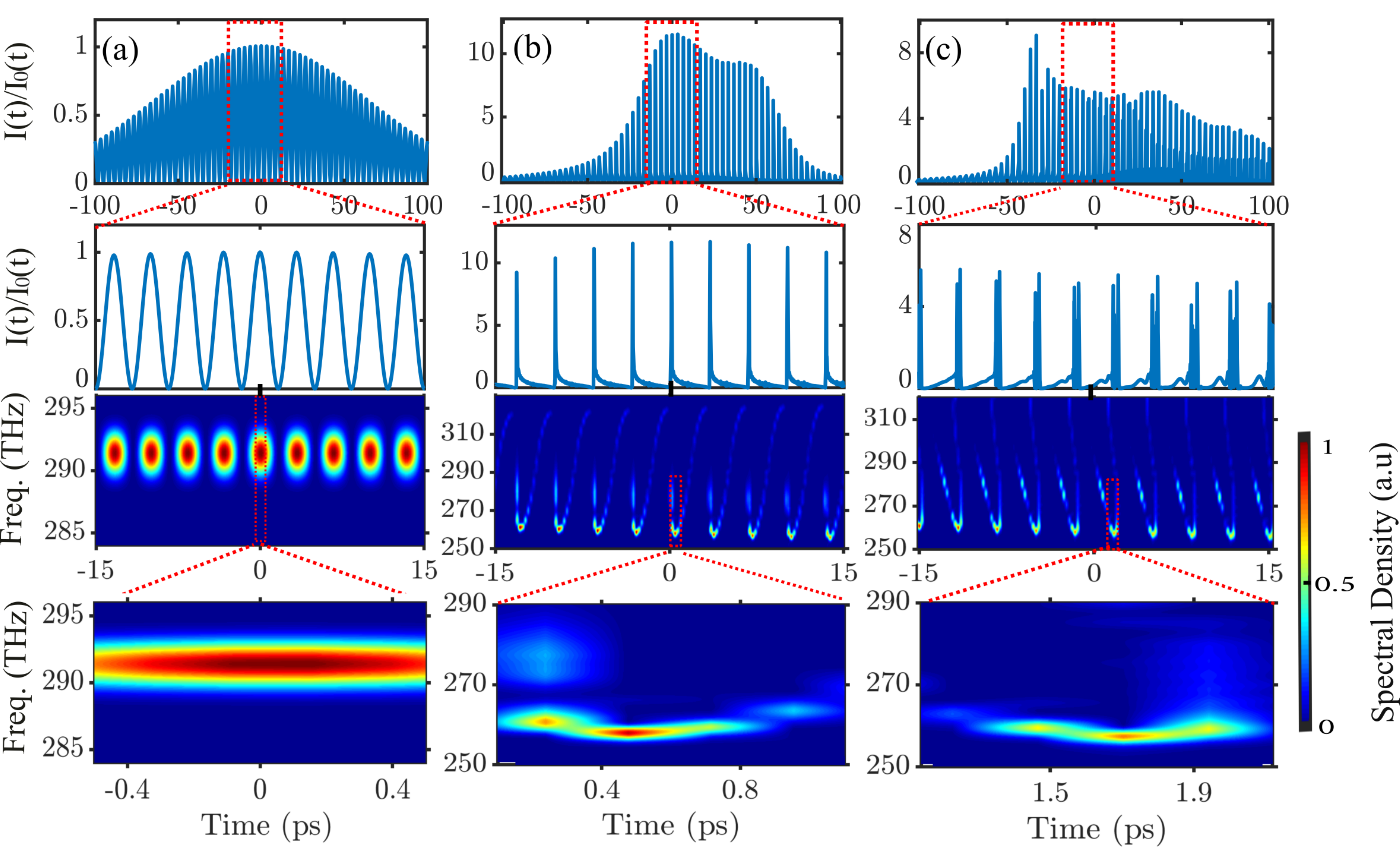}} 
 
\caption{(a) shows the input pump pulse, (b) shows the output pump pulse after one $2.5\,\text{cm}$ PPLN stage. (c) shows the output pump pulse after dispersion compensation by adding the opposite second order dispersion of Lithium Niobate at the pump center frequency (GDD$=6300$\,$\mathrm{fs^2}$) to (b). The contour plot in the second and third rows are the short time Fourier transforms of the corresponding selected temporal range (indicated by the dotted red box). \label{2dtf} }
\end{figure}

\begin{figure}[H]
  \centering

{\includegraphics[width=13.6cm]{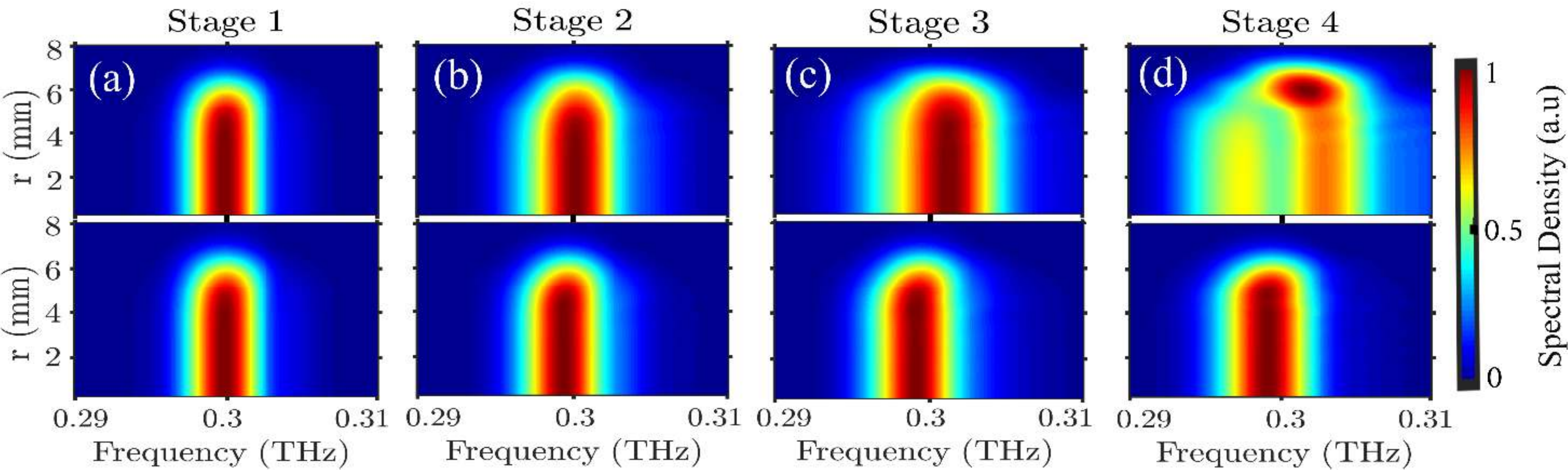}} 
 
\caption{(a-d) show the terahertz spectral density generated by 4 consecutive stages respectively. The first and second rows show the terahertz spectra generated without and with dispersion compensation, respectively. \label{4_stages_eff_thz}}
\end{figure}

\begin{figure}[H]
  \centering

{\includegraphics[width=13.6cm]{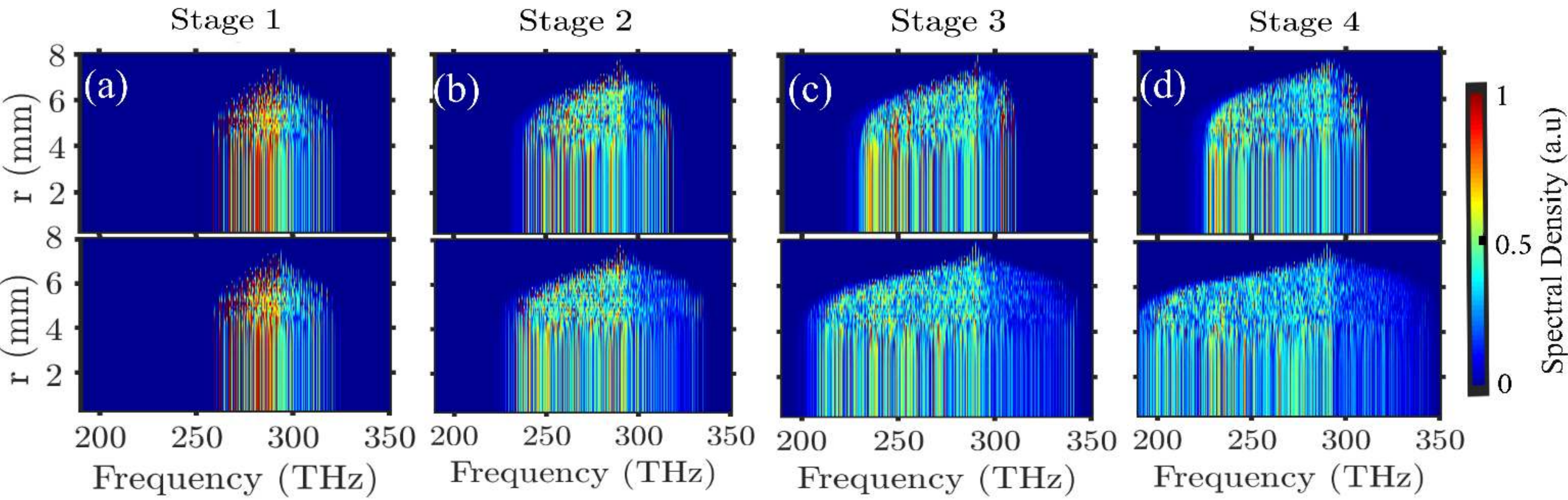}} 
 
\caption{(a-d) represent pump pulse spectral density after each corresponding stage. The first and second rows show the pump pulse spectra at the output of each stage generated without and with dispersion compensation, respectively.\label{4_stages_eff_ir}}
\end{figure}

Figure (\ref{2dtf}) depicts the temporal profile of the pump pulse with the short time Fourier transform (STFT) (i.e instantaneous spectrum) of the corresponding time range. 
The broadening of the pump pulse spectrum causes a reduction of the pulse duration, an increase in peak intensity and a drastic variation in the instantaneous spectrum of each sub-pulse (see Fig.\,\ref{2dtf}{(a,b)}).  The STFT in Fig.\,\ref{2dtf}{(b)} indicates that the instantaneous spectrum in each sub-pulse forms a 'U' shape due to the cascading effect, leading to a drastic spectral variation in time. The maximum cascading (spectral down-shift) occurs where the highest peak intensity is present. The STFT in Fig.\,\ref{2dtf}{(c)} suggests that the dispersion compensation alters the spectral distribution with respect to time. If a transform limited pump pulse, which has a uniform spectrum distributed over time, could be obtained after each stage, the terahertz efficiency could be greatly enhanced. However, the realization of such a process is challenging, since perfect compression of the 'U' shaped spectrum induced by SPM, the second order nonlinear effect, and dispersion, is not straightforward. 
Figure.\,\ref{4_stages_eff_thz} compares the spectral density (spatially dependent spectra in radial dimension $r$) of the terahertz generated at each stage for two cases with and without dispersion compensation. The terahertz beams generated with dispersion compensation follow a Gaussian-like spectrum, whereas those generated by direct pump pulse recycling have distorted spectral shapes, particularly in the last stage. This is due to the fact that the phase of the pump perturbed by dispersion influences the phase-matching condition of terahertz generation, and thus leads to a broader terahertz spectrum. Figure.\,\ref{4_stages_eff_ir} shows the comparison of the spectral density of the output optical pump at each stage without and with dispersion compensation. It can be seen that dispersion compensation boosts the cascading process and leads to greater pump spectral broadening. The results depicted in Fig.\,\ref{4_stages_eff1}, Fig.\,\ref{4_stages_eff_thz} and Fig.\,\ref{4_stages_eff_ir} confirm that the terahertz conversion efficiency can be largely enhanced by pump pulse recycling with dispersion compensation. 
 
\subsubsection{Quartz Output Coupler for Beam Separation }\label{my_coupler}

\begin{figure}[H]
  \centering

{\includegraphics[width=10.3cm]{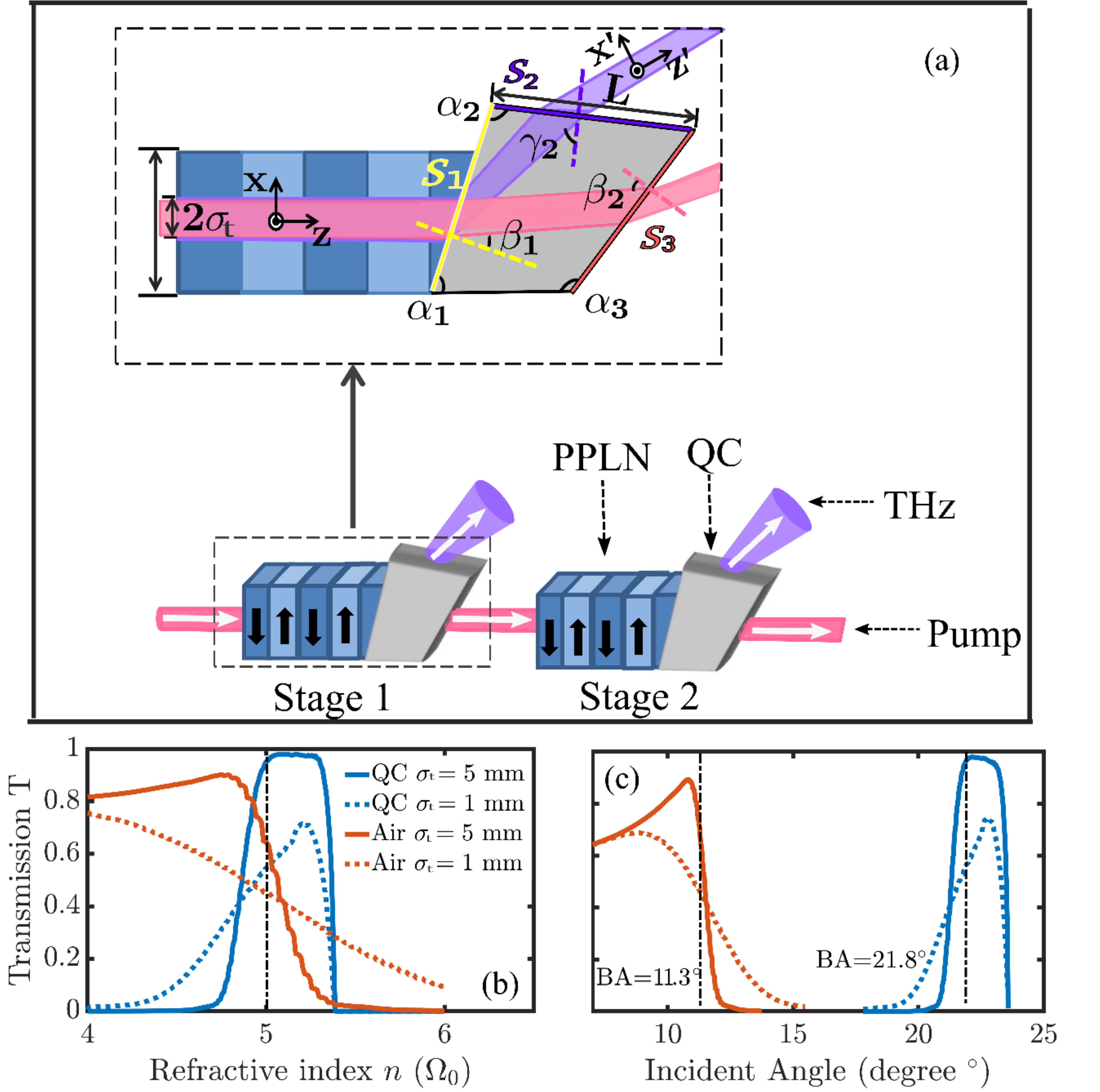}} 
 
\caption{(a) Schematic illustration of the output coupler and the angle definitions. In (b) and (c), comparison of the terahertz energy transmission from PPLN to air for 'Air' and 'QC' cases are shown. Additionally, the transmission of the generated flat-top terahertz beam with different beam sizes $\sigma_\text{t}=1$\,mm (dashed lines) and $\sigma_\text{t}=5$\,mm  (solid lines) are analyzed. (b) shows terahertz energy transmission versus the refractive index of the PPLN at the given Brewster angles (BA) at $S_1$ surface.  The black dash-dotted line refers to the refractive index assumed for the calculation of Brewster's angles. (c) shows terahertz energy transmission versus the incident angle at surface $S_1$ with given refractive index where the black dash-dotted lines mark the BA at this refractive index.\label{brewster_angel1} }
\end{figure} 
Figure.\,\ref{brewster_angel1}{(a)} shows one possible arrangement of the multi-stage setup accounting for the subtlety of terahertz and optical beam separation. The output pump beam from the PPLN propagates at a certain angle with respect to the input pump pulse direction due to the refraction from the quartz coupler (QC). However, in the sketch, we do not depict this for simplicity of illustration. Incidence at Brewster angle plays a crucial role in this scheme in order to couple out and separate both pump and terahertz beams with high transmission efficiency from the PPLN to air ($n=1$).
Out-coupling of the terahertz beam from Lithium Niobate to air through a silicon coupler was first demonstrated experimentally and theoretically by Kawase et al.  \cite{kawase1997unidirectional}.
In this article, crystalline quartz is chosen as the material of the output coupler owing to its low absorption for terahertz radiation at $0.3\,$THz (0.01/cm) \cite{naftaly2007terahertz,grischkowsky1990far}, large bandgap (12.5 eV) \cite{ghosh1999dispersion} which has lower nonlinearity and higher damage threshold compared to silicon or Lithium Niobate\cite{shoji1997absolute}. Additionally, the dispersion of quartz at both pump and terahertz frequencies is negligible for few centimeters of propagation \cite{grischkowsky1990far,ghosh1999dispersion}. We neglect the small influence of the quartz coupler on the optical pump and fully focus on the out-coupling of the terahertz radiation.

In Fig.\,\ref{brewster_angel1}{(a)}, the inner angles of the bulk quartz coupler are represented by $\alpha_1$, $\alpha_2$, and $\alpha_3$. This ensures that the terahertz beams experience Brewster incidence angles $\pi-\alpha_1$ and $\gamma_2$ at $S_1$ and $S_2$ surfaces, respectively. $\beta_1$ and $\beta_2$ are the output angles of the pump beam at the $S_1$ surface and the pump beam Brewster incidence angle at the $S_3$ surface, respectively. The refraction loss of the pump is negligible. $S_3$ can be chosen to be parallel to surface $S_1$ to further reduce angular deviation. Additionally,  $L$ (the length of the top edge of the quartz coupler) is the minimum distance that ensures sufficient spatial separation of pump and terahertz beams. Details of the equations used for calculating the involved angles can be found in Table\,\ref{parameter_table2}, where  $\sigma=5$\,mm, $n_\text{LN}(\omega_0)=2.1$, $n_\text{Q}(\omega_0)=1.4,n_\text{LN}(\Omega_0)=5$, $n_\text{Q}(\Omega_0)=2$. The subscripts 'LN' and 'Q' correspond to PPLN and Quartz respectively.

\bgroup
\def\arraystretch{1.3}
\begin{table}[H]
\centering \caption{Quartz Coupler Parameters \label{parameter_table2} }
\begin{tabular}{|c| c| c| }
\hline
\textbf{Parameter}& \textbf{Variable} & \textbf{Values}\\
\hline
 $\beta_{1}$ & $\sin^{-1}\left(\cos(\alpha_1)n_{\text{LN}}(\omega_0)/n_\text{Q}(\omega_0)\right)$&$33.9\degree$\\ 
\hline
   $\beta_{2}$ &  $\tan^{-1}(1/{n_\text{Q}(\omega_0)})$&$35.5 \degree$  \\
\hline

   $\gamma_2$ &  $\tan^{-1}(1/{n_\text{Q}(\Omega_0)})$&$26.6\degree$ \\
\hline
   $\alpha_1$  & $\tan^{-1}\left( n_\text{LN}(\Omega_0)/n_\text{Q}(\Omega_0) \right)$&$68.2 \degree$  \\
  
\hline

  $\alpha_2$& $\alpha_{1}+\gamma_{2}$&$94.8\degree$ \\
\hline
$\alpha_3$ & $\pi-\beta_{1}+\beta_{2}-\alpha_{1}$&$113.4\degree$\\
\hline
$L$&$2\sigma \cos(\beta_{2})\cos(\alpha_1)/(\sin(\alpha_1-\beta_1)\sin(\alpha_2))$&$5.5$ mm\\
\hline

\end{tabular}
\end{table}
\egroup
Since the refractive index of the PPLN varies with temperature and manufacturing process, an output coupler with a reasonable tolerance is desired. In Fig.\,\ref{brewster_angel1}{(b,c)}, comparison of two out-coupling cases 'Air' (red curves) and 'QC' (blue curves) is shown. In the case of 'Air' where no output coupler is utilized, the PPLN crystal is cut such that the terahertz pulse is incident at Brewster's angle ($\tan^{-1}(1/{n_\text{LN}(\Omega_0)})=\tan^{-1}(1/{5})=11.3 ^{\circ}$) from PPLN to air directly. In the case of 'QC', the terahertz pulse is coupled out from the PPLN to air through the designed quartz coupler shown in Fig.\,\ref{brewster_angel1}{(a)} with Brewster incidence angle ($\tan^{-1}(n_\text{Q}{(\Omega_0)}/{n_\text{LN}(\Omega_0)}=\tan^{-1}(2/{5})=21.8 ^{\circ}$) on $S_1$ surface.  

In both Fig.\,\ref{brewster_angel1}{(b)} and Fig.\,\ref{brewster_angel1}{(c)}, the transmission window of the 'QC' case at $\sigma_\text{t}=5\,$mm (flat interval with terahertz energy transmission $T\approx 1$) is due to total internal reflection (TIR) at both the $S_1$ and $S_2$ surfaces. This window range reduces for the smaller terahertz beam size $\sigma_\text{t}=1\,$mm (i.e terahertz beam with larger angular divergence). 
The terahertz beam components with angular divergence larger than the TIR are filtered out, leading to lower energy transmission. This TIR induced transmission loss is particularly pronounced for the 'Air' case since the refractive index varies from $n_\text{LN}(\Omega_0)=5$ to $n_\text{Air}(\Omega_0)=1$. This large refractive index difference causes a sharp transmission drop at the incidence angle larger than Brewster angle. This explains why the transmission peak occurs at an angle smaller than Brewster's angle in the 'Air' case in Fig.\,\ref{brewster_angel1}{(c)}. With larger beam size $\sigma_\text{t}=5$\,mm, the transmission peak approaches the Brewster angle due to smaller angular divergence. 
It can be seen that from PPLN to air, the terahertz energy transmission at $\sigma_\text{t}=1$\,mm is around $45 \%$ and $55\%$ for 'Air' case and 'QC' case respectively. The terahertz energy transmission at $\sigma_\text{t}=5$\,mm  is around $63 \%$ and $96\%$ for 'Air' case and 'QC' case  respectively. With the given refractive index at the corresponding Brewster incident angle, a quartz coupler leads to higher transmission. More discussion can be found in section\,(\ref{set_up_sketch}).

\section{Analysis}

\subsection{Single Stage: Terahertz Efficiency in Terms of Pump Pulse Duration}\label{tau_vs_eff}

\begin{figure}[H]
  \centering
  
{\includegraphics[width=12cm]{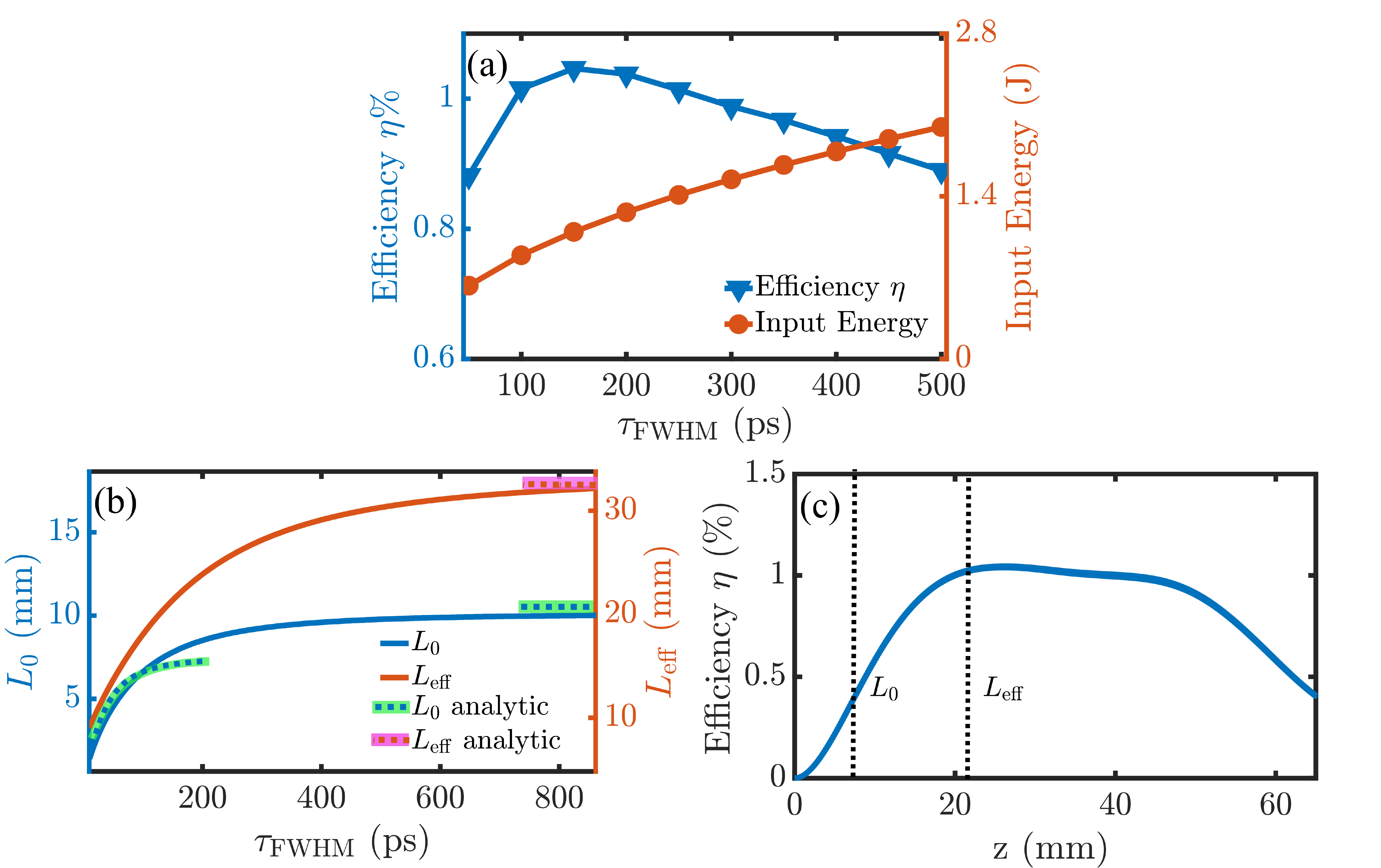}}

\caption{Numerical results of the terahertz generation in a single stage: (a) shows the conversion efficiency versus the pump pulse duration. The maximum efficiency $\eta=1.05~\%$ is obtained at $\tau_{\text{FWHM}}=150$\,ps. The input pump energy is calculated from Eq.\,(\ref{energy_input}) with the pump beam size $\sigma=5$\,mm. (b) shows the $L_\text{0}$, $L_{\text{eff}}$ versus pump pulse duration. (c) shows the terahertz efficiency versus the propagation distance for $\tau_{\text{FWHM}}=150$\,ps, where the steepest efficiency increase rate ($L_\text{0}$) and effective length ($L_{\text{eff}}$) are labeled.    \label{scan_tau}  }
\end{figure} 
The crucial parameters in designing a complete terahertz source are the pulse duration of the pump and the necessary length of the PPLN crystal. As one can see in Eq.\,(\ref{energy_input}), longer pulse durations allow higher input energy, but simultaneously necessitate longer interaction length, which in turn intensifies the effects caused by terahertz absorption. The optimal pump pulse duration is found to be $150\,$ps with conversion efficiency $\eta=1.05~\%$ as shown in Fig.\,\ref{scan_tau}{(a,c)}. 
In order to define the optimal PPLN length $L_{\text{eff}}$, we introduce the length parameter $L_0$ where the efficiency increase is most rapid. This is obtained at $d^2 \eta(z)/dz^2|_{z=L_0}=0$. We define $L_{\text{eff}}$ as:

\begin{equation}\label{leff_define}
\frac{1}{e}\dfrac{d\eta(z)}{dz}|_{z=L_0}=\dfrac{d\eta(z)}{dz}|_{z=L_\mathrm{eff}}.
\end{equation}
One should avoid choosing $L_{\text{eff}}$ at $d\eta(z)/dz=0$, because the plateau length in Fig.\,\ref{scan_tau}{(c)} increases with the increase of the pump pulse duration and $d\eta(z)/dz=0$ occurs approximately in the middle of the plateau. Within the range of the plateau, the amount of generated terahertz radiation equals to the absorbed amount. Consequently, the optimal $L_{\text{eff}}$ should be chosen at the onset of the plateau to avoid unnecessarily long crystals.

If we define $\delta \approx  \left[{n(\Omega_{0})-n_\mathrm{g}(\omega_0)}\right]/({c \tau_{\text{FWHM}}})$, Eq.\,(\ref{leff_define}) results in the following analytic equations for $L_{\text{eff}}$ and $L_0$ (see Appendix 
\ref{tau_L} for details):
\begin{equation}
\begin{cases}
\text{for short pump pulses}, \frac{\delta}{\alpha} \gg 1, L_0=\tan^{-1}(\frac{\delta}{\alpha})/{\delta}\\
\text{for long pump pulses}, \frac{\delta}{\alpha} \ll 1, L_0={2\ln{(2)}}/{\alpha},\,L_{\text{eff}}=\frac{2}{\alpha}\ln{(\frac{2}{1-\sqrt{1-e^{-1}}})}
\end{cases} \label{l_eff_calculate}
\end{equation}
$L_\text{0}$ and $L_{\text{eff}}$ variations in terms of the pump pulse duration are depicted in Fig.\,\ref{scan_tau}{(b)} where the results of analytic calculations are compared against numerical results.

\subsection{Single Stage: Effects of Transverse Spatial Variations of Pump}
High power terahertz generation demands high power lasers which usually suffer from phase front distortions and intensity modulations. In the following section, we investigate the influence of the spatial variation of the pump pulse on the terahertz generation process. 
\subsubsection{Beam Size Dependence}\label{section_beam_size}

\begin{figure}[H]
  \centering
  
{\includegraphics[width=12cm]{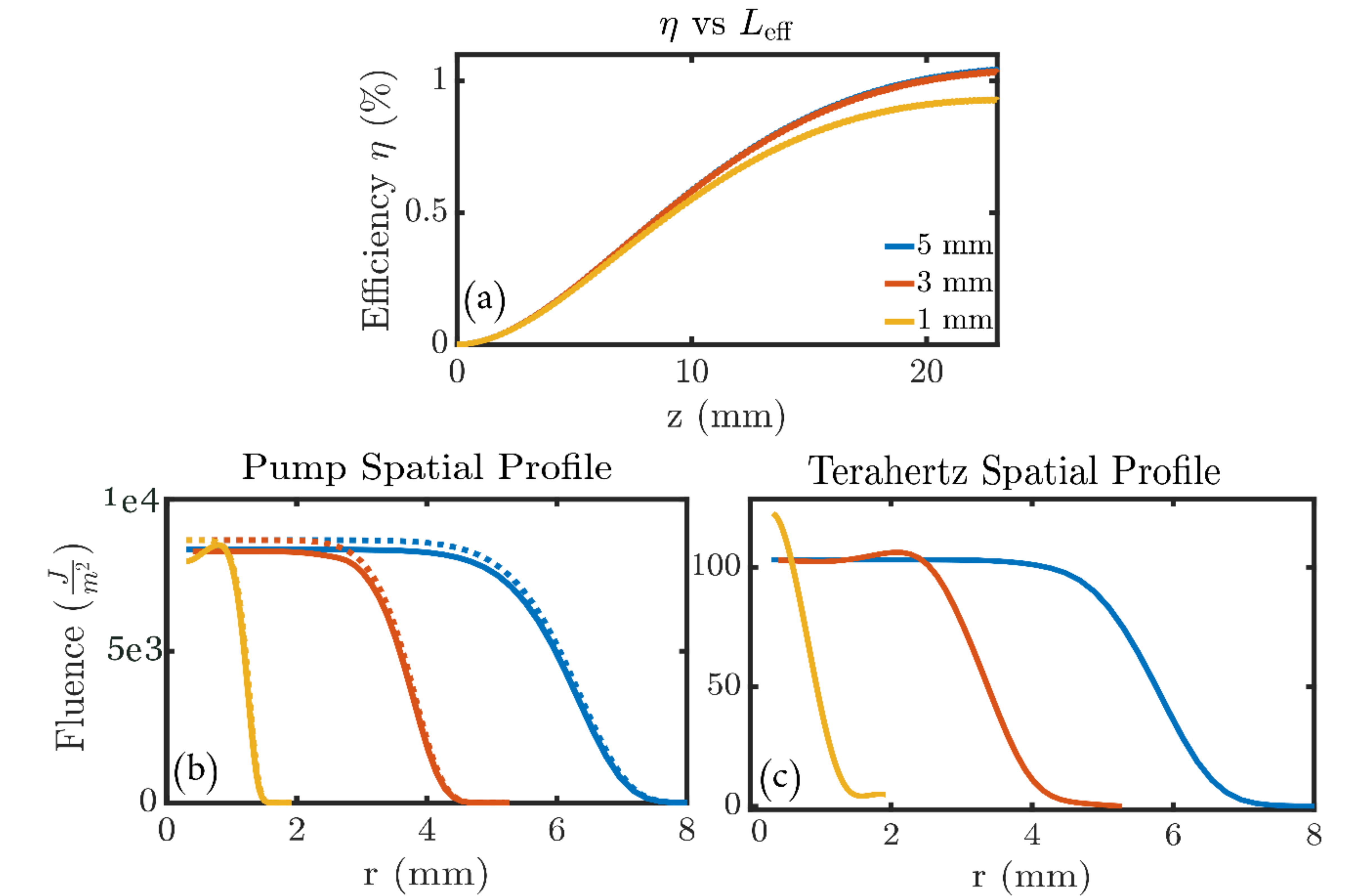}}

\caption{(a) shows the efficiency of terahertz generation versus propagation distance for various beam sizes. (b) shows the spatial profile of the input pump pulse (dashed lines) and output pump pulse (solid lines). (c) shows the output terahertz beam spatial profile. For different beam sizes, the peak fluence is kept constant. The output spatial profiles of both terahertz radiation and the pump are obtained at the end of the interaction length, i.e. $z=25$\,mm.\label{compare_spatial} }
\end{figure}

Here, we aim at exploring the dependence of the terahertz generation efficiency on the pump beam size.
For this purpose, we perform simulations for a super-Gaussian ($M=5$) pump beam with beam sizes $\sigma=1$\,mm, $3$\,mm and $5$\,mm, respectively. Figure\,\ref{compare_spatial} suggests that the terahertz beam size and spatial profile resemble the corresponding pump properties for large pump beam sizes.

It can be seen in Fig.\,\ref{compare_spatial}{(a)} that the terahertz conversion efficiency is lower when $\sigma=1$\,mm compared to $\sigma=3$\,mm and $5$\,mm. The diffraction has a minor influence on the spatial profile of the pump pulse since the Rayleigh range ($\lambda \approx 1$\,\textmu m) is about few tens of meters. In a super-Gaussian beam, smaller beam size leads to larger spatial phase gradient at the edge of the spatial profile. Therefore, as observed in Fig.\,\ref{compare_spatial}{(b)}, the self-focusing effect leads to the peak appearing at the edge of the 1 mm pump beam.
The change of the spatial profiles of the terahertz beams in Fig.\,\ref{compare_spatial}{(c)} is due to the diffraction of the super-Gaussian beam \cite{parent1992propagation,chafiq2007flat}. It can be seen that for a 1 mm terahertz beam, diffraction modifies the super-Gaussian profile to a Gaussian profile after 25 mm generation distance.

The above results are obtained with perfect super-Gaussian pump spatial profiles and a flat spatial phase front. In order to examine the influence of spatial energy and phase front variations of the pump pulse on terahertz generation with the two-line pump spectrum, we studied 4 cases: (\rom{1}) flat phase front
with one of the pump spectral lines having a $5\% $ modulation in the electric field strength. (\rom{2}) flat phase front
with both pump spectral lines having a $5\% $ modulation in the electric field strength. (\rom{3}) flat spatial profile with one of the pump spectral lines having modulation from $-\pi$ to $\pi$ in the phase front. (\rom{4}) flat spatial profile with both of the pump spectrum lines having modulation from $-\pi$ to $\pi$ in phase front. 

In the following context, $E(\omega,r)$ denotes the electric field of the input pump pulse and $E_0(\omega,r)$ is the electric field with a perfect super-Gaussian spatial profile ($\sigma$=5\,mm) and flat phase front. The interaction length is chosen to be $z=25$\,mm as in Fig.\,\ref{compare_spatial}. $r_{\text{max}}=8.5\,$mm is the maximum distance of the calculated radial dimension.

\subsubsection*{Case \rom{1}: One line with Cosine Electric Field Amplitude Modulation}
\begin{equation}\label{2lines_inten}
\begin{cases}
E(\omega,r)=E_0(\omega,r)(1+0.05\cos(2\pi\times\frac{4r}{r_{\text{max}}})),  \quad \omega\leq2\pi\times 291.3\text{\,THz}\\
E(\omega,r)=E_0(\omega,r),  \quad \omega>2\pi\times 291.3\text{\,THz}
\end{cases}
\end{equation}

\begin{figure}[H]
  \centering
  
{\includegraphics[width=9.7cm]{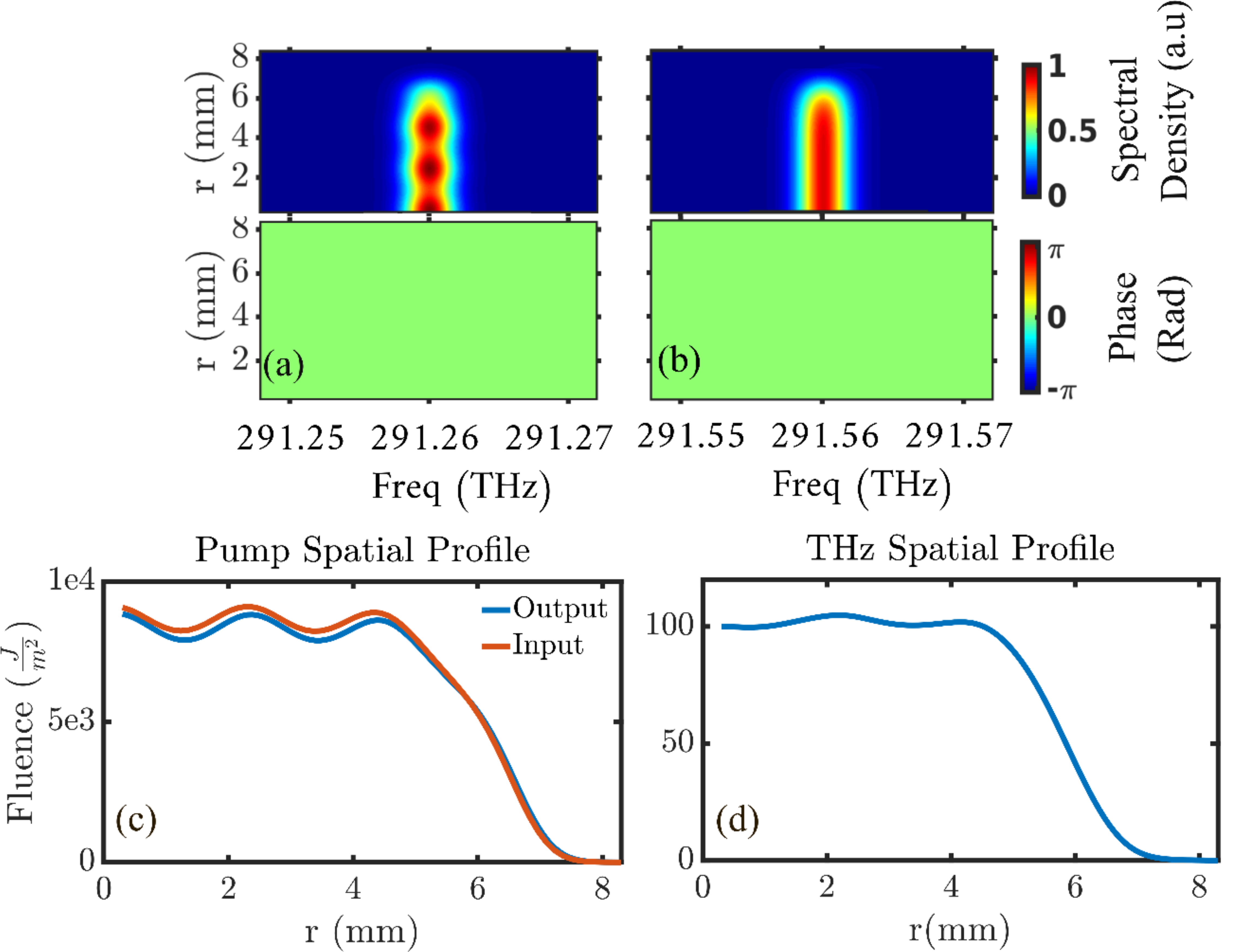}}

\caption{(a) and (b) show the spectral density (first row) and flat phase front (second row) of the input pump spectral lines. (c) shows the spatial profile of the input and output pump pulse. (d) shows the terahertz spatial profile. \label{case_1} }
\end{figure}

It can be seen in Fig.\,\ref{case_1} that terahertz generation is not degraded by the intensity modulation in one of the spectral lines. Though the transverse intensity modulation can also induce spatial phase variations due to self-focusing, with our pump pulse parameters, the induced spatial phase variation is negligible. 

\subsubsection*{ Case \rom{2}: Both lines with Cosine Electric Field Amplitude Modulation }
\begin{equation}
E(\omega,r)=E_0(\omega,r)(1+0.05\cos(2\pi\times\frac{4r}{r_\mathrm{max}}))
\end{equation}

\begin{figure}[H]
  \centering
  
{\includegraphics[width=10cm]{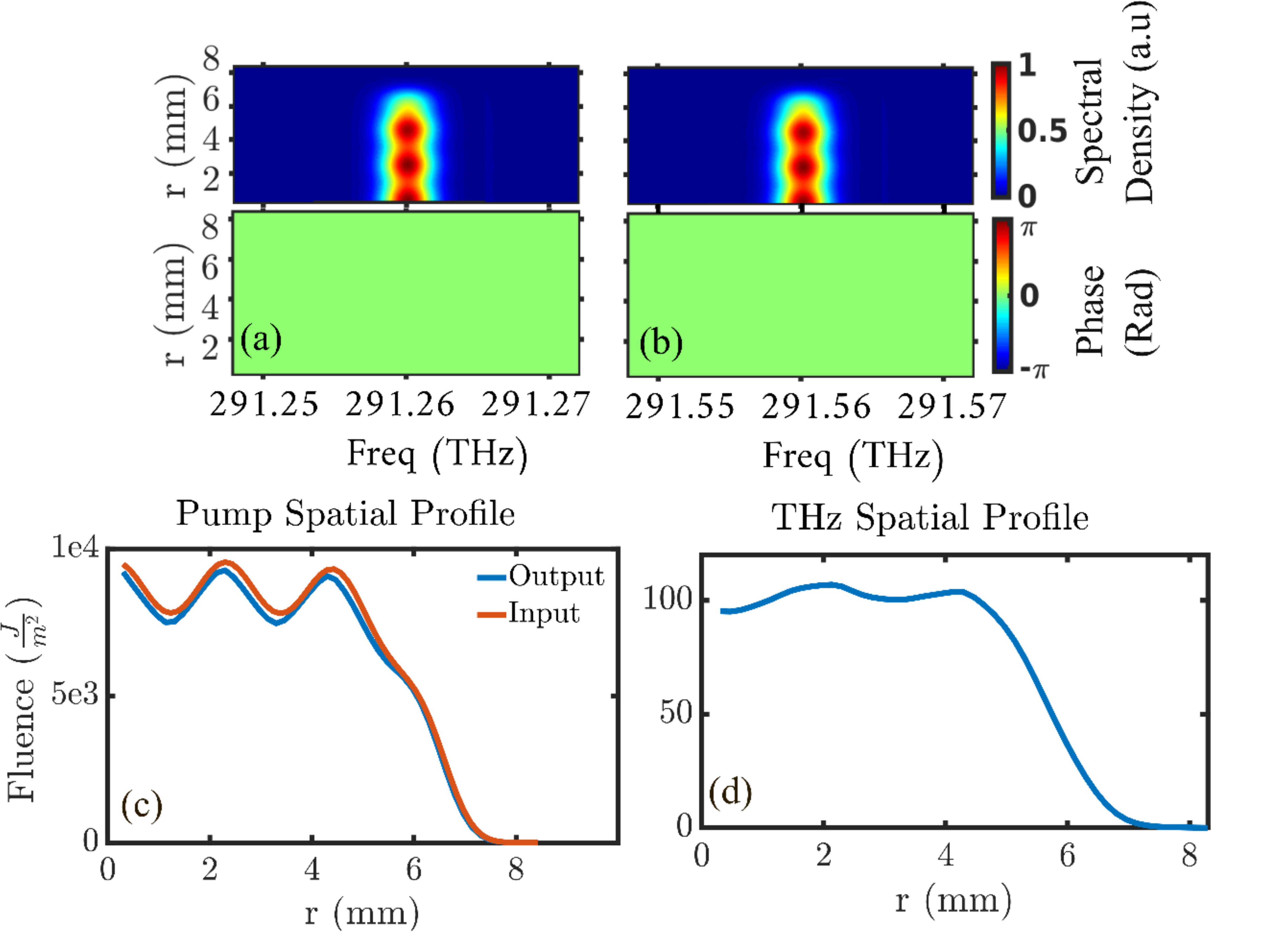}}

\caption{(a) and (b) show the spectral density (first row) and flat phase front (second row) of the input pump spectral lines. (c) shows the spatial profile of the input and output pump pulse. (d) shows the terahertz spatial profile.\label{case_2} }
\end{figure}

 Fig.\,\ref{case_2} suggests that the terahertz generation is not degraded when both lines are subjected to the same intensity modulation, which is expected from the Case I study.

\begin{figure}[H]
  \centering
{\includegraphics[width=10cm]{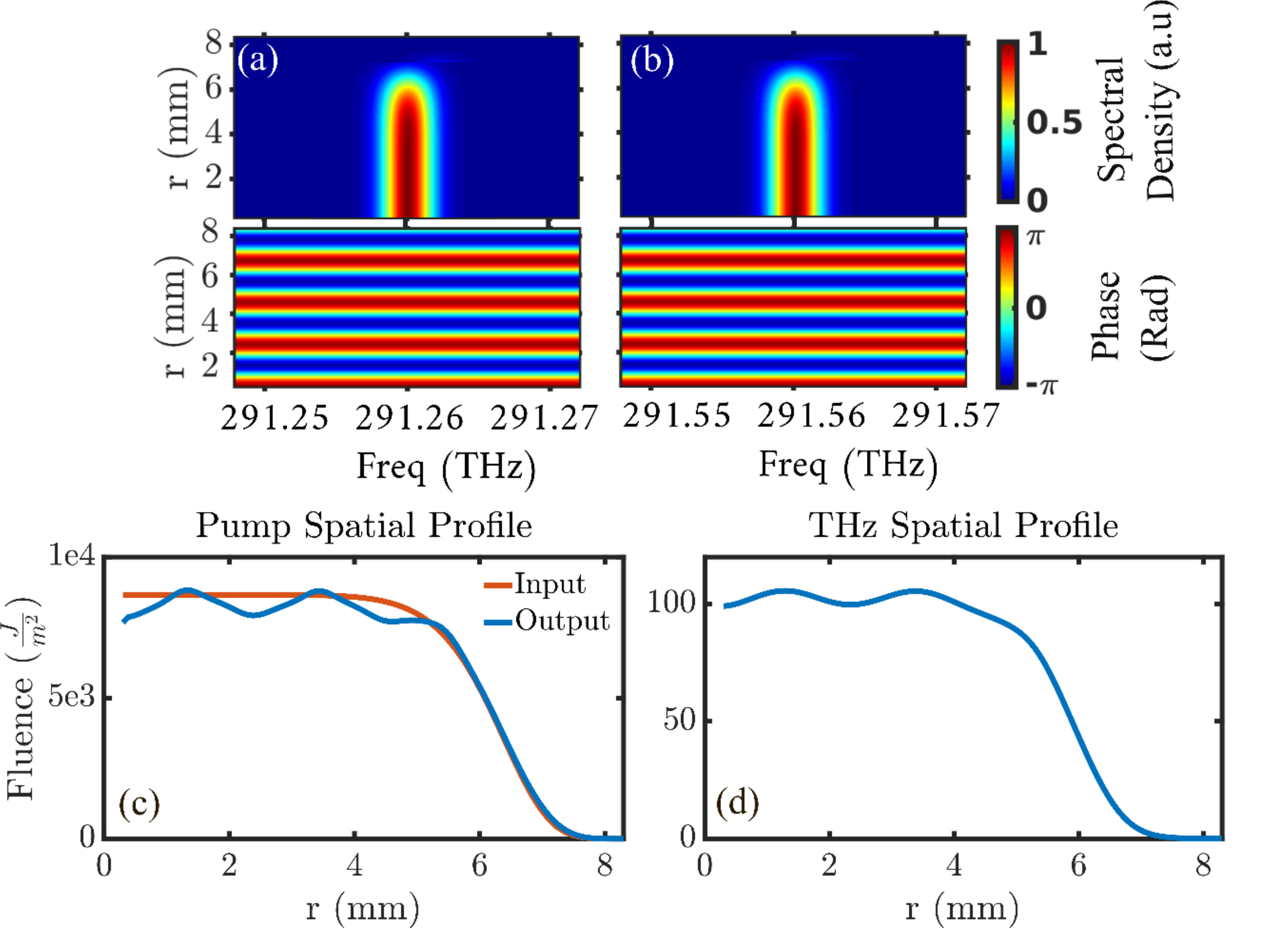}}

\caption{(a) and (b) show the spectral density (first row) and phase front (second row) of the input pump spectral lines. (c) shows the spatial profile of the input and output pump pulse. (d) shows the terahertz spatial profile.\label{case_4}  }
\end{figure}

\subsubsection*{Case \rom{3}: Both lines with Cosine Phase Modulation }

\begin{equation}
E(\omega,r)=E_0(\omega,r)e^{i\pi\cos(2\pi\times\frac{4r}{r_{\text{max}}})}
\end{equation}
Fig.\,\ref{case_4} suggests that if both of the pump lines are identically modulated in phase, the terahertz generation process is not degraded. The phase modulation induces self-focusing which can be observed in Fig.\,\ref{case_4}{(c)}.

\subsubsection*{Case \rom{4}: One line with Cosine Phase Modulation}
\begin{figure}[H]
  \centering
{\includegraphics[width=9.5cm]{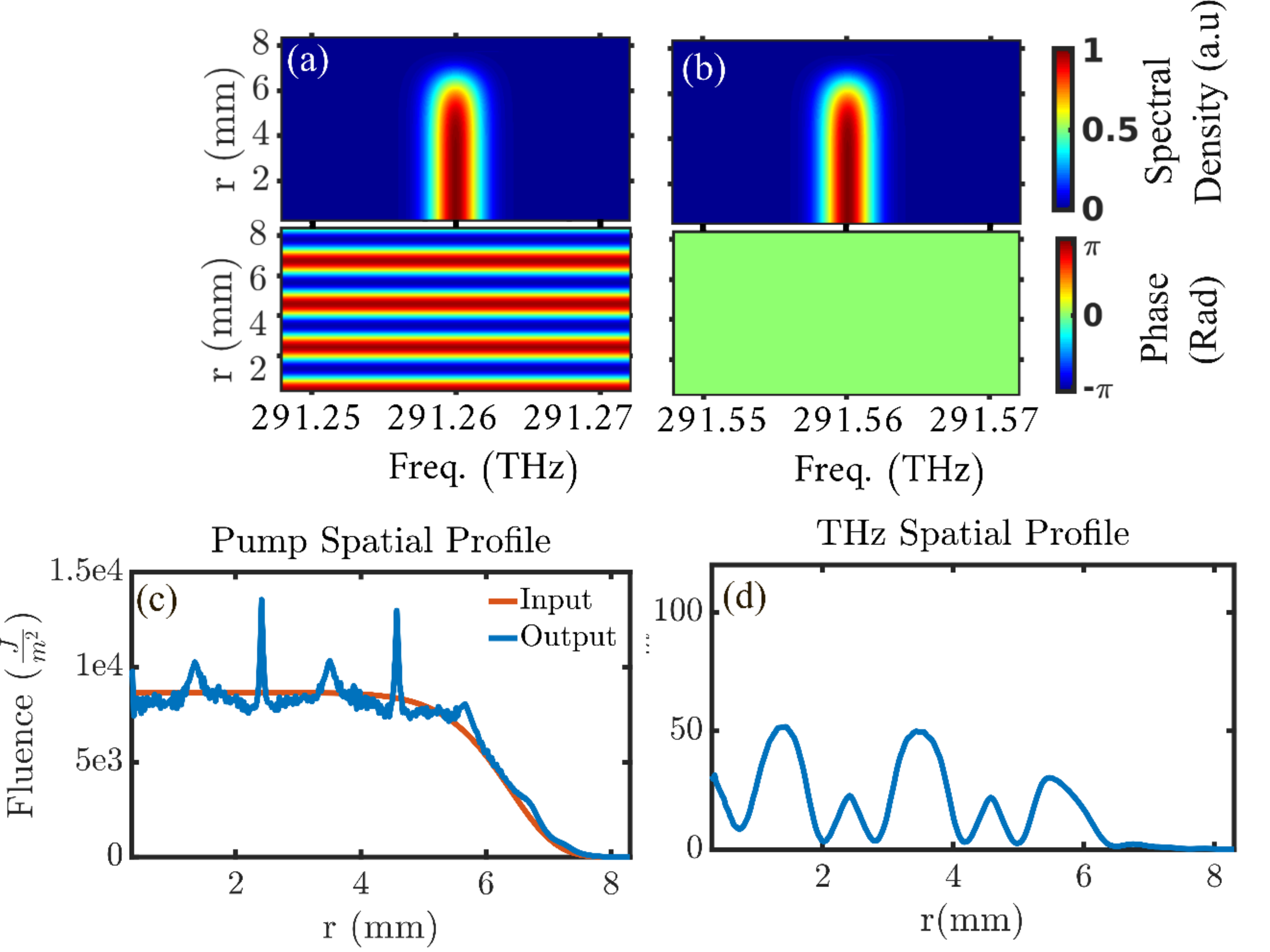}}

\caption{In (a,b) the first row shows the spatial profiles of the pump spectral lines and the second row represents the phase front of the corresponding spectral lines. (c) shows the pump input and output spatial profile. (d) shows the generated terahertz spatial profile.\label{case_3}}
\end{figure}

\begin{equation}\label{one_sin_phase}
\begin{cases}
E(\omega,r)=E_0(\omega,r)e^{i\pi\cos(2\pi\times\frac{4r}{r_{\text{max}}})},  \quad \omega\leq2\pi\times 291.3\text{\,THz}\\
E(\omega,r)=E_0(\omega,r),  \quad \omega>2\pi\times 291.3\text{\,THz}
\end{cases}
\end{equation}
Figure\,\ref{case_3} indicates that a distortion in the phase front, not unusual for high power laser beams, strongly reduces the conversion efficiency. This is due to the nature of the difference frequency generation process, which transfers the phase difference between the two frequency lines of the optical pump to the terahertz beam at a thousand times longer wavelength. In turn, the modulated terahertz phase front suffers from strong diffraction, leading to destructive interference. Consequently, the terahertz conversion efficiency is greatly reduced and a strongly spatially modulated terahertz beam appears as shown in Fig.\,\ref{case_3}{(d)}. The peaks in Fig.\,\ref{case_3}{(c)} are due to self-focusing, making it different from case \rom{3}. In case \rom{3}, since the phase fronts of both spectral lines are identical, the terahertz generated inherits a flat phase front. As a result, the cascaded frequencies newly generated in the optical region maintain the same phase front as the initial two-line spectra. However, in case \rom{4} since DFG creates a terahertz pulse with the phase difference of the pump spectral lines, the new pump frequency lines generated by the cascading process carry different phases. The later generated cascaded lines inherit multiple times of the original phase difference, forming an enormously curved phase front. Thus, strong self-focusing occurs.

\subsection{Terahertz Spatial Profile after Quartz Coupler} \label{set_up_sketch}
\begin{figure}[H]
  \centering

{\includegraphics[width=13.6cm]{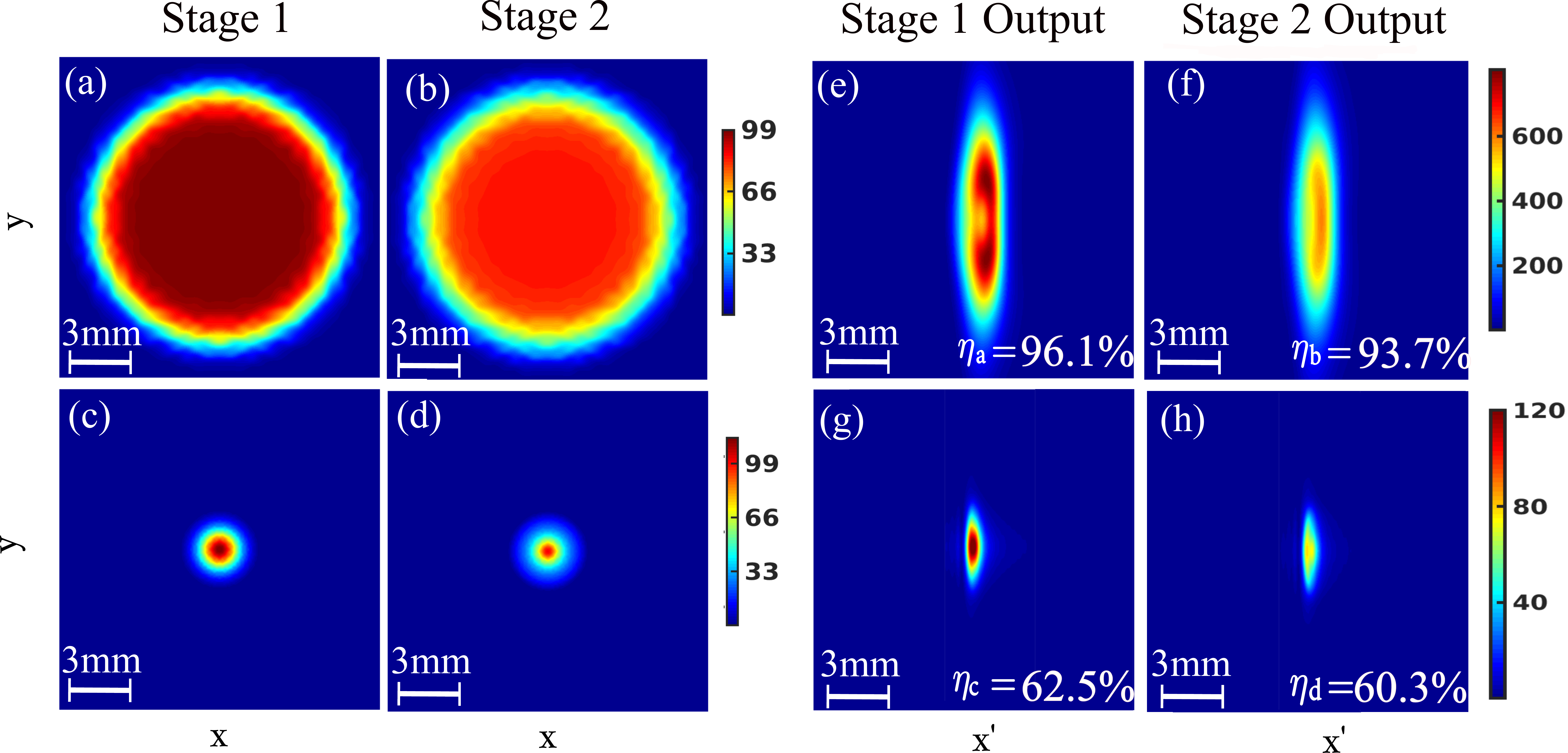}} 
 
\caption{ (a,b) and (c,d) represent terahetz spatial profiles in PPLN in $x-y$ coordinate before $S_1$ surface (see Fig.\,\ref{brewster_angel1}{(a)}) generated by a two-stage system with pump beam size $\sigma=5$\,mm and $\sigma=1$\,mm respectively.  (e-h) represent coupled out terahertz spatial profiles of  (a-d) after $S_2$ surface in $x'-y$ coordinate in the air. The color bar represents fluence in the unite of J/m$^2$. Each figure window size is 15\, mm $\times$ 15\,mm. Terahertz out-coupling  efficiency from (a-d) to (e-h) is $\eta_{\text{a}}=96.1\%,\eta_{\text{b}}=93.7\%,\eta_{\text{c}}=62.5\%,\eta_{\text{d}}=60.3\%$ respectively. \label{cartisin_spatial} }
\end{figure} 
 The quartz coupler breaks the cylindrical symmetry of the PPLN. In order to calculate the terahertz spatial profile after the quartz coupler, we reconstruct the terahertz beam profiles in x-y coordinates from the numerical results in the cylindrical coordinates inside the PPLN before the incidence on the $S_1$ surface (see Fig.\,\ref{brewster_angel1}{(a)}).
The refraction at the $S_1,S_2$ surfaces (see Fig.\,\ref{brewster_angel1}{(a)}) for p polarized light and the beam propagation in the quartz coupler is calculated by the angular spectrum method \cite{goodman2005introduction}. Figs.\,\ref{cartisin_spatial}{(a,b)} and Figs.\,\ref{cartisin_spatial}{(c,d)} correspond to terahertz spatial profiles generated with a pump beam size $\sigma=5$\,mm and $\sigma=1\,$mm, respectively by a two-stage setup.

 Note that with a pump beam size $\sigma=5$\,mm, the terahertz spatial profile retains its flat top nature, whereas with $\sigma=1$\,mm, the terahertz spatial profile reduces to a Gaussian as shown in Fig.\,(\ref{compare_spatial}). The corresponding transmitted spatial profiles after the $S_2$ surface in $x'-y$ coordinates in the air (see Fig.\,\ref{brewster_angel1}{(a)}) are shown in Fig.\,\ref{cartisin_spatial}{(e-h)} with the terahertz out-coupling  efficiency $\eta_{\text{a}}=96.1\%,\eta_{\text{b}}=93.7\%,\eta_{\text{c}}=62.5\%,\eta_{\text{d}}=60.3\%$ respectively.
The influence of the $S_1$ and $S_3$ surfaces together with the propagation in the quartz coupler has a negligible effect on the pump spatial profile, i.e the spatial profile of the pump stays flat top but with a reduced size in the $x'$ direction i.e $\sigma'=\sigma\tan(\beta_2)\cos(\beta_1)/ \sin(\alpha_1) \approx 0.6 \sigma$. The beam size of the terahertz radiation in the $x'$ direction follows the same relation $\sigma'_\mathrm{t}=\sigma_\mathrm{t}\tan(\beta_2)/{\tan(\alpha_1)}\approx 0.2 \sigma_\mathrm{t}$. 
On the other hand, the refraction at $S_1$ and $S_2$ surfaces greatly changes the energy distribution of the flat top terahertz beams spatially ( see Fig.\,\ref{cartisin_spatial}{(a,b,e,f)} ) while the Gaussian spatial profile remains unchanged ( see Fig.\,\ref{cartisin_spatial}{(c,d,g,h)} ). The flat top function is a superposition of Hermite Gaussian modes in Cartesian coordinates.
Thus, the final spatial profile of terahertz is a result of a superposition of all the Hermite Gaussian modes after refraction.

The terahertz beams coupled out by the QC could be combined by a power combiner with a specifically designed input coupler which matches the terahertz beam profile to the mode of the power combiner \cite{landy2009guiding}. This investigation is beyond the scope of this article and will be presented in a separate study. For an ideal symmetric power combiner, the combination efficiency $\eta_\text{combine}$ can be written as the following \cite{gupta1992power}: 

\begin{equation}\label{combine_eta}
\eta_{\text{combine}}=\frac{\int_{-\infty}^{\infty}\int_{-\infty}^{\infty}|\sum_{k=1}^{N_\mathrm{T}} E_\mathrm{k}(x,y,t)|^2 dxdydt}{N_\mathrm{T}\int_{-\infty}^{\infty}\int_{-\infty}^{\infty}\sum_{k=1}^{N_\mathrm{T}} |E_\mathrm{k}(x,y,t)|^2 dxdydt}
\end{equation}
where $E_\mathrm{k}(x,y,t)$ is the terahertz electric field obtained from $k$th stage.
The combination efficiency $\eta_{\text{combine}}$  reaches the maximum when $E_1(x,y,t)=E_2(x,y,t)=.....=E_\mathrm{N}(x,y,t)$, e.g terahertz outputs from each stage need to have identical spectrum, phase and amplitude \cite{gupta1992power}. This could be achieved by adjusting the length of each stage and the relative arrival time of the terahertz beams. 
With equation (\ref{combine_eta}), the combination efficiency of the terahertz output beams are $93.0\%$ (Fig.\,\ref{cartisin_spatial}{(e,f)}) and $95.2\%$ ( Fig.\,\ref{cartisin_spatial}{(g,h)}), respectively. This high combination efficiency is the result of similarity between terahertz beams generated in each stage. As a result, the final total efficiency $\eta_{\text{total}}$ of a two stage system can be written in Eq.\,(\ref{eff_total}) where $\eta_1,\eta_2$ is the generation efficiency of the first and second stage respectively, $\eta_{\text{a}},\eta_{\text{b}}$ is the out-coupling efficiency, from PPLN to the air with QC, of the first and second stage respectively and $\eta_{\text{combine}}$ is the combination efficiency of the terahertz generated by the first and the second stage.  
 
\begin{equation}\label{eff_total}
\eta_{\text{total}}=\eta_{\text{combine}}\times(\eta_{1}\times\eta_{\text{a}}+\eta_{2}\times\eta_{\text{b}})
\end{equation}
Using equation (\ref{eff_total}) and (\ref{energy_input}), a total terahertz energy $17.6$\,mJ with $\eta_{\text{total}}=1.6\%$ can be achieved with a two stage system using pump beams of size $\sigma=5\,\text{mm}$ and input pump energy $1.1$\,J.

\section{ Conclusion}
We presented a scheme for generating terahertz pulses using a consecutive arrangement of PPLN crystals (multi-stage system). For this purpose, a 2-D numerical scheme assuming cylindrical symmetry is developed. The simulation includes the effects of DFG, SPM, self-focusing, beam diffraction, dispersion and terahertz absorption. 

  For a single stage, the optimal pump beam pulse duration for a 2-line pump spectrum configuration is $150$\,ps with effective length $25$\,mm and conversion efficiency $\eta=1.05\%$. With this pump pulse configuration, the terahertz spatial profile resembles the pump when the pump waist $\sigma\geq3\,$ mm. Additionally, we found that the intensity profile modulation has negligible influence on the terahertz generation efficiency, whereas a distortion in phase front strongly reduces the conversion efficiency. This suggests that care must be taken in circumventing phase induced efficiency deterioration for high energy terahertz generation.
  
For a multi-stage system, we found that by recycling the pump pulse with dispersion compensation, the efficiency can be greatly enhanced. Subsequently, a quartz output coupler for separating the terahertz beam and optical pump with high terahertz transmission is designed.

Specifically, for a two stage system, we predict generation of a terahertz pulse of $17.6$ mJ energy with total conversion efficiency $\eta_\mathrm{total}=1.6\%$ at $0.3$\,THz for a 1.1\,J input pump energy, where the generation efficiencies are $\eta_{1}=1.0\%,\eta_{2}=0.8\%$, the out-coupling efficiencies are $\eta_\mathrm{a}=96.1\%,\eta_\mathrm{b}=93.7\%$ and combination efficiency is $\eta_\mathrm{combine}=93.0\%$.

\section*{Acknowledgments}

In addition to DESY, this work has been supported by the European Research Council under the
European Union’s Seventh Framework Programme (FP7/2007-2013) through the Synergy Grant
AXSIS (609920), the excellence cluster ‘The Hamburg Center for Ultrafast Imaging – Structure,
Dynamics and Control of Matter at the Atomic Scale’ (CUI, DFG-EXC1074). We thank the
DESY HPC team for providing access and computation time on the DESY HPC cluster. Lu Wang would like to thank IMPRS (International Max Planck Research School for Ultrafast Imaging $\&$ Structural Dynamics). The authors additionally would like to thank Dr. Nicholas Matlis and Dr. Steve Aplin for the supportive and helpful discussions.
 
\section{Appendix}
\subsection{Phase-matching}  \label{pm_appendix}
In PPLN, the pump pulse is chosen to be linearly polarized along the extraordinary optical axis of LiNbO\textsubscript{3} in order to maximize the effective second order nonlinear coefficient $d_{33}=\frac{1}{2} \chi^{(2)}_{333}=-\frac{1}{4}n^4(\omega)\gamma_{333}=168 \, $pm/V \cite{weis1985lithium,vodopyanov2006optical,hebling2008generation,ravi2016pulse}. 
\begin{equation}\label{chi_2}
\begin{bmatrix}
P^{NL_{(2)}}_1\\
P^{NL_{(2)}}_2\\
P^{NL_{(2)}}_3\\

\end{bmatrix}
=2\epsilon_0
\begingroup
\setlength\arraycolsep{2pt}
\begin{bmatrix}
   0 & 0 & 0 & 0 & d_{15}&-d_{22} \\
   -d_{22}  & d_{22} & 0 & d_{15} &0 &0\\
   d_{15}  & d_{15} & d_{33} &0  &0 &0
\end{bmatrix}
\endgroup
\begingroup
\setlength\arraycolsep{2pt}
\begin{bmatrix}
|E_1|^2\\
|E_2|^2\\
|E_3|^2\\
E_2E_3^*+E_2^*E_3\\
E_1E_3^*+E_1^*E_3\\
E_1E_2^*+E_1^*E_2
\end{bmatrix}
\endgroup
\end{equation}
In Eq.\,(\ref{chi_2}), subscripts 1 and 2 represent the ordinary axes and 3 represents the extraordinary axis \cite{boyd2003nonlinear}. Axes 1 and 2 are identical at the first order ($n_\text{o1}=n_\text{o2}$), which is known as uniaxial birefringence. However, in the second order, the o axes are not identical. This explains why $d_{ij}$ matrix elements are not identical on 1 and 2 axes. With the linearly polarized input pump pulse $ \hat{\bf{E}}=(0,0,E_3) $ along the extraordinary axis, the second order polarization term driving the terahertz field is also linearly polarized along the same direction ($P^\mathrm{NL_{(2)}}_3=2\epsilon_0d_{33}|E_3|^2$).

The phase-matching condition shown in Eq.\,(\ref{pm_ppln}) ensures that the terahertz pulses generated at different positions of the PPLN crystal can add up in phase provided:
\begin{eqnarray}
&&\Omega\large[{1}/{v_\mathrm{THz}(\Omega)}-{1}/{v_\mathrm{g}(\omega)}\large]\frac{\Lambda }{2}=\pi+2\pi N \label{pm_ppln}\\
&&N=0,1,2,3....\nonumber
\end{eqnarray}

where $\Lambda$ is the period of PPLN, $\Omega$ is the terahertz angular frequency, $\omega$ is the pump angular frequency, $v_\mathrm{THz}$ is the phase velocity of terahertz pulse and $v_g$ is the group velocity of the pump pulse. Note that for a given PPLN structure, there is more than one terahertz frequency fulfilling the phase-matching condition ( $\Omega_N = \Omega_0 \times(1 + 2N)$) 
\subsection{Frequency Dependent Self Phase Modulation}\label{f_spm}
Frequency dependent SPM includes self-steepening which can be shown using the following derivation:
\begin{eqnarray}
&&\omega_0\mathscr{F} \Bigg \lbrace |A_\mathrm{op}(t,r,z)|^2A_\mathrm{op}(t,r,z)-\frac{i}{\omega_0}\frac{\partial|A_\mathrm{op}(t,r,z)|^2A_\mathrm{op}(t,r,z)}{\partial t} \Bigg \rbrace\\
&&=\omega_0\mathscr{F}\lbrace |A_\mathrm{op}(t,r,z)|^2A_\mathrm{op}(t,r,z)\rbrace+(\omega-\omega_0)\mathscr{F}\lbrace|A_\mathrm{op}(t,r,z)|^2A_\mathrm{op}(t,r,z)\rbrace \nonumber \\
&&=\omega\mathscr{F}\lbrace|A_p(t,r,z)|^2A_\mathrm{op}(t,r,z)\rbrace. \nonumber 
\end{eqnarray}
where the first and second terms correspond to the SPM effect at $\omega_0$ and self-steepening, respectively.
\subsection{ Finite Difference Method}\label{numerical_method_ap}
The finite difference methodology for solving Eq.\,(\ref{ir_cylin}) is presented. Equation (\ref{thz_cylin}) can be solved by following the same procedure. By defining  $r^+_{j}={\frac{1}{2}(r_{j+1}+r_{j})}/{(\bigtriangleup r^2 r_{j}}), \ r^-_{j}={\frac{1}{2}(r_{j-1}+r_{j})}/({\bigtriangleup r^2 r_{j}})$, where $\bigtriangleup r$ is the step size in the $r$ dimension and $r_j$ represents a mesh point at a specific radial position, one can obtain the discretized form of the second order derivative as shown in Eq.\,(\ref{discretization_r}). 
\begin{align}
\frac{1}{r_j}\frac{\partial}{\partial r}\left({r_j\frac{\partial A_\mathrm{op}(\omega_m,r_j,z_k)}{\partial r}}\right)&=
r^-_{j}A_\mathrm{op}(\omega_m,r_{j-1},z_k)-(r^-_{j}+r^+_{j})A_\mathrm{op}(\omega_m,r_j,z_k)\nonumber\\  
&+ r^+_{j}A_\mathrm{op}(\omega_m,r_{j+1},z_k)\label{discretization_r}
\end{align}
Equation (\ref{discretization_r}) can be written in a matrix multiplication form as shown on the right hand side of the last column in Eq.\,(\ref{matrix_f}) where the $r_i$ related matrix is denoted as a tridiagonal matrix $\mathbf{M_R}$ as shown in Eq. (\ref{r_matrix}).\\
\begin{minipage}[t]{0.1\textwidth}
\vspace{2.4cm}
$\bf{M_R}=$
\end{minipage}
\begin{minipage}[t]{0.9\textwidth}
\begin{eqnarray*}\label{r_matrix}
    \begin{tikzpicture}
        \matrix [matrix of math nodes,nodes={minimum width=3.em, inner sep=3.1pt},left delimiter=(,right delimiter=)] (m)
        {
     -r^{-}_1& r^+_{1}+r^-_{1}& -r^+_{1} & 0 &0 & \dots  &0 \\
    0 & -r^-_{2} & r^+_{2}+r^-_{2} & -r^+_{2} &0& \dots  &0 \\
    0 &  0 &-r^-_{3} & r^+_{3}+r^-_{3} & -r^+_{3} & \dots  &0 \\
    \vdots&\vdots & \vdots & \vdots & \ddots & \vdots &0 \\
    0&0 & 0 & \dots &  -r_{N_r}^{-}  &  r^+_{N_r}+r^-_{N_r} &-r^{+}_{N_{r}+1}        \\
        };  
\draw[<->,transform canvas={xshift=-1.5em},thick] (m-5-1.south west) -- node[left=2pt] {$N_r$} (m-1-1.north west);

\draw[<->,transform canvas={yshift=1.2em},thick] (m-1-1.north west) -- node[above=2pt] {$N_r+2$} (3.7,1.65);

\draw[color=black , dashed] (m-1-1.north west) -- (m-1-1.north east) -- (m-5-1.south east) -- (m-5-1.south west) -- (m-1-1.north west);
 
 \draw[color=black , dashed]   ([xshift=-17.5pt,yshift=8pt]m-1-7.center) -- ([xshift=17.5pt,yshift=8pt]m-1-7.center) -- (m-5-7.south east) -- (m-5-7.south west) -- ([xshift=-17.5pt,yshift=8pt]m-1-7.center);
    \end{tikzpicture}  
\end{eqnarray*}
\end{minipage}
\begin{minipage}[t]{0.03\textwidth}

\begin{equation}
\end{equation}
\end{minipage}\\
Equation (\ref{ir_cylin}) can be written as Eq.\,(\ref{finite_difference_r})
where $P^\mathrm{NL}(\omega_m,r_j,z_k)$ represents all the second and third nonlinear polarization terms. 

 \begin{align}
\frac{\partial A_\mathrm{op}(\omega_m,r_j,z_k)}{\partial z}&=P^\mathrm{NL}(\omega_m,r_j,z_k)+\frac{i}{2k(\omega_m)}[-r^-_{j}A_\mathrm{op}(\omega_m,r_{j-1},z_k)\nonumber\\  
&+(r^-_{j}+r^+_{j})A_\mathrm{op}(\omega_m,r_j,z_k)- r^+_{j}A_\mathrm{op}(\omega_m,r_{j+1},z_k)]\label{finite_difference_r}
\end{align}

Equation (\ref{finite_difference_r}) can be written in matrix form as Eq.\,(\ref{matrix_f}) where $N_r,N_\omega$ are the number of the mesh grids at $r$ and $\omega$ dimension respectively. Each color of the corresponding column vector represents the coupled wave equation at a specific frequency $\omega_m$. In Eq.\,(\ref{matrix_f}), the 2-D matrix of angular frequency ($\omega$) and radius ($r$) on the left hand side is updated by a low-storage Runge-Kutta method \cite{williamson1980low}. The electric field elements at the same position $r_j$ with different frequencies are mixed by the nonlinear polarization $P^\mathrm{NL}$. The electric field elements of the same frequency $\omega_m$ at different positions in $r$ are mixed by $\mathbf{M_R}$ (e.g diffraction). The totally number of the mesh points in $r$ dimension is $N_r+2$. The extra 2 points $A_\mathrm{op}(\omega,r_0,z),A_\mathrm{op}(\omega,r_{N_{r}+1},z)$, marked with dashed boxes, are "ghost points". The "ghost points" together with the two columns marked by dashed lines in $\mathbf{M_R}$ in Eq.\,\ref{r_matrix} are used to construct boundary conditions.\\

\begin{minipage}{.3\textwidth}
\begin{tikzpicture}[every node/.style={anchor=north east,minimum width=1mm,minimum height=7.65mm}]
\matrix (mA) [draw, text width=2.05cm,fill=blue!10,matrix of math nodes]
{
 \frac{\partial A_\mathrm{op}(\omega_{N_\omega},r_1,z_k)}{\partial z}  \\
 \frac{\partial A_\mathrm{op}(\omega_{N_\omega},r_{2},z_k)}{\partial z}  \\
\vdots\\
 \frac{\partial A_\mathrm{op}(\omega_{N_\omega},r_{N_r},z_k)}{\partial z}  \\
};

\matrix (mB) [draw,text width=1.95cm,fill=yellow!10,matrix of math nodes] at ($(mA.south west)+(2.1,0.2)$)
{
 \frac{\partial A_\mathrm{op}(\omega_m,r_1,z_k)}{\partial z}  \\
 \frac{\partial A_\mathrm{op}(\omega_m,r_{2},z_k)}{\partial z}  \\
\vdots \\
 \frac{\partial A_\mathrm{op}(\omega_m,r_{N_r},z_k)}{\partial z}  \\
};

\matrix (mC) [draw,text width=1.95cm,fill=red!10,matrix of math nodes] at ($(mB.south west)+(2.1,0.2)$)
{
 \frac{\partial A_\mathrm{op}(\omega_1,r_1,z_k)}{\partial z}  \\
 \frac{\partial A_\mathrm{op}(\omega_1,r_{2},z_k)}{\partial z}  \\
\vdots \\
 \frac{\partial A_\mathrm{op}(\omega_1,r_{N_r},z_k)}{\partial z}  \\
} node [right,xshift=-0.7cm,yshift=-9cm,font=\bf] {\LARGE=};

\draw[<-,thick,black] ($(mA.north west)+(-0.15,0)$) -- ($(mC.north west)+(-0.15,0)$) node [pos=0.66,left,xshift=0.0cm,yshift=2cm,rotate=80] {Frequency ($\omega$)};
\draw[->,thick,black] ($(mC.north west)+(-0.15,-0.1)$) -- ($(mC.south west)+(-0.15,0)$)
node [left,xshift=-0.25cm,yshift=2.5cm,rotate=90] {Radius ($r$)};
\draw[dashed](mA.north east)--(mC.north east);
\draw[dashed](mA.north west)--(mC.north west);
\draw[dashed](mA.south east)--(mC.south east);

\end{tikzpicture}
\end{minipage}
\begin{minipage}{.25\textwidth}
\begin{tikzpicture}[every node/.style={anchor=north east,text width=24mm,minimum height=6.72mm}]
 \matrix (mA) [draw,minimum width=7mm,fill=blue!10,matrix of math nodes]
{
P^\mathrm{NL}(\omega_{N_{\omega}},r_1,z_k) \\
P^\mathrm{NL}(\omega_{N_{\omega}},r_2,z_k) \\
\vdots \\
P^\mathrm{NL}(\omega_{N_{\omega}},r_{N_r},z_k) \\
};

\matrix (mB) [draw,text width=1.95cm,fill=yellow!10,matrix of math nodes] at ($(mA.south west)+(2.55,-0.14)$)
{
P^\mathrm{NL}(\omega_{m},r_1,z_k) \\
P^\mathrm{NL}(\omega_{m},r_2,z_k) \\
\vdots \\
P^\mathrm{NL}(\omega_{m},r_{N_r},z_k) \\
};

\matrix (mC) [draw,text width=1.95cm,fill=red!10,matrix of math nodes] at ($(mB.south west)+(2.55,-0.14)$)
{
P^\mathrm{NL}(\omega_{1},r_1,z_k) \\
P^\mathrm{NL}(\omega_{1},r_2,z_k) \\
\vdots \\
P^\mathrm{NL}(\omega_{1},r_{N_r},z_k) \\
}node [right,xshift=-0.3cm,yshift=-8.8cm,font=\bf] {\LARGE+};

\draw[dashed](mA.north east)--(mC.north east);
\draw[dashed](mA.north west)--(mC.north west);
\draw[dashed](mA.south east)--(mC.south east);

\end{tikzpicture}
\end{minipage}
\begin{minipage}{.25\textwidth}

\begin{tikzpicture}[every node/.style={anchor=north east,minimum width=1.4cm,minimum height=4.3mm}]

 \matrix(mA) [draw,text width=2.65cm,fill=blue!10,matrix of math nodes,nodes={minimum width=7em, inner sep=1pt},      
        b/.style={fill=white}] at ($(mA.north east)+(2.3,0.7)$)
{
|[b]| A_\mathrm{op}(\omega_{N_{\omega}},r_0,z_k) \\
A_\mathrm{op}(\omega_{N_{\omega}},r_1,z_k) \\
A_\mathrm{op}(\omega_{N_{\omega}},r_2,z_k) \\
\vdots \\
A_\mathrm{op}(\omega_{N_{\omega}},r_{N_r},z_k) \\
|[b]| A_\mathrm{op}(\omega_{N_{\omega}},r_{N_r+1},z_k) \\
}node [left,xshift=-0.7cm,yshift=-1cm] {$\frac{i}{2k(\omega_{N_{\omega}})} \bf{M_R}\times$};

 \draw[color=black,dashed] (mA-1-1.north west) -- (mA-1-1.north east) -- (mA-1-1.south east) -- (mA-1-1.south west) -- (mA-1-1.north west);
 \draw[color=black,dashed] (mA-6-1.north west) -- (mA-6-1.north east) -- (mA-6-1.south east) -- (mA-6-1.south west) -- (mA-6-1.north west);

\matrix(mB)  [draw,fill=yellow!10,matrix of math nodes,nodes={minimum width=2.1em, inner sep=1pt},      
        b/.style={fill=white}] at ($(mA.south west)+(2.7,-0.2)$)
{
|[b]| A_\mathrm{op}(\omega_{m},r_0,z_k) \\
A_\mathrm{op}(\omega_{m},r_1,z_k) \\
A_\mathrm{op}(\omega_{m},r_2,z_k) \\
\vdots \\
A_\mathrm{op}(\omega_{m},r_{N_r},z_k) \\
|[b]| A_\mathrm{op}(\omega_{m},r_{N_r+1},z_k) \\
}node [left,xshift=-1.05cm,yshift=-4.5cm] {$\frac{i}{2k(\omega_m)} \bf{M_R}\times$};

 \draw[color=black,dashed] (mB-1-1.north west) -- (mB-1-1.north east) -- (mB-1-1.south east) -- (mB-1-1.south west) -- (mB-1-1.north west);
 \draw[color=black,dashed] (mB-6-1.north west) -- (mB-6-1.north east) -- (mB-6-1.south east) -- (mB-6-1.south west) -- (mB-6-1.north west);

\matrix (mC) [draw,fill=red!10,matrix of math nodes, 
        nodes={minimum width=2.1em, inner sep=1pt},      
        b/.style={fill=white}] at ($(mB.south west)+(2.7,-0.2)$)
{ |[b]|  A_\mathrm{op}(\omega_{1},r_0,z_k)  \\
A_\mathrm{op}(\omega_{1},r_1,z_k) \\
A_\mathrm{op}(\omega_{1},r_2,z_k) \\
\vdots \\
A_\mathrm{op}(\omega_{1},r_{N_r},z_k) \\
|[b]| A_\mathrm{op}(\omega_{1},r_{N_r+1},z_k) \\
}node [left,xshift=-1.23cm,yshift=-7.5cm] {$\frac{i}{2k(\omega_1)} \bf{M_R}\times$};

 \draw[color=black,dashed] (mC-1-1.north west) -- (mC-1-1.north east) -- (mC-1-1.south east) -- (mC-1-1.south west) -- (mC-1-1.north west);
 \draw[color=black,dashed] (mC-6-1.north west) -- (mC-6-1.north east) -- (mC-6-1.south east) -- (mC-6-1.south west) -- (mC-6-1.north west);

\draw[dashed](mA.north east)--(mC.north east);
\draw[dashed](mA.north west)--(mC.north west);
\draw[dashed](mA.south east)--(mC.south east);

\end{tikzpicture}
\end{minipage}
\begin{equation}\label{matrix_f}
\end{equation}

\subsection{ Boundary Conditions}
The boundary conditions are enforced to the solution by the elements in the dashed boxes in Eq.\,(\ref{matrix_f}), and $\bf{M_R}$ matrix in Eq.\,(\ref{r_matrix}). For the mesh points in $r$ dimension, the $r=0$ singularity is avoided by defining $r_0=3\Delta r$, and thus  $r^-_{1}=\frac{1}{2}(r_{0}+r_{1})/({\bigtriangleup r^2 r_{1}})$. The "boundary condition" near $r_1$ is confined by the symmetry at the origin of the cylindrical coordinate itself ( e.g $\frac{\partial A_\mathrm{op}(\omega,r,z)}{\partial r}|_{r=r_1}=0$). Thus,
\begin{equation}
A_\mathrm{op}(\omega,r_0,z)=A_\mathrm{op}(\omega,r_{1},z) 
\end{equation}
For the boundary condition at $r_{N_r}$, three different conditions can be applied. Note that as shown in Fig.\,(\ref{cylin_slice}), in the calculated area the polarization of the field is perpendicular to the boundary. Of course with the same numerical method, any polarization at the boundary can be considered. However, the boundary condition has to be changed accordingly.
\begin{enumerate}
\item Dirichlet boundary.\\
This condition represents the interface from PPLN crystal ($n_\mathrm{op}$) to air ($n$). Due to Gauss's law in maxwell equation $\iint	
\overrightarrow{ \mathbf{D}}\cdot\mathbf{d}\overrightarrow{\mathbf{S}}	=0$. One can get
\begin{eqnarray}
A_\mathrm{op}(\omega,r_{N_{r}+1},z)=A_\mathrm{op}(\omega,r_{N_r},z)\frac{n_\mathrm{op}^2r_{N_r}  }{n^2r_{N_{r}+1}}
\end{eqnarray}

If the field polarization is parallel to the contact surface of the dielectric (PPLN) and the metal cladding outside the PPLN, one can also apply a reflective boundary condition. 
\begin{equation}
A_\mathrm{op}(\omega,r_{N_{r}+1},z)=0.
\end{equation}
This is equivalent to the situation where electric field is reflected by a mirror.
\item Transparent boundary \cite{hadley1992transparent}.\\
This boundary represents the situation where the boundary has no effect on the field as the field propagates through the calculation  boundary. This kind of boundary is also equivalent to the situation when Eq. (\ref{ir_cylin}) and (\ref{thz_cylin}) are solved by angular spectrum method. The condition can be written as \cite{hadley1992transparent}.
\begin{eqnarray}
A_\mathrm{op}(\omega,r_{N_{r}+1},z_{k+1})\approx A_\mathrm{op}(\omega,r_{N_r},z_{k+1})\frac{A_\mathrm{op}(\omega,r_{N_{r}+1},z_{k})r_{N_r}}{A_\mathrm{op}(\omega,r_{N_r},z_{k})r_{N_{r}+1}}
\end{eqnarray}
\end{enumerate}

\subsection{Pulse Duration in Terms of Effective Length }\label{tau_L}
As section (\ref{tau_vs_eff}) suggests, the effective length is related to the pump pulse duration. In order to get more insight into this problem, we begin with a 1-D undepleted model. Consequently, Eq.\,(\ref{thz_cylin}) can be written as Eq.\,(\ref{1d_thz}).
\begin{eqnarray}
A_\mathrm{THz}(\Omega,z)=&&\frac{-i\Omega \chi^{(2)}_{333} }{\pi n(\Omega) c }\int_{0}^\infty A_\mathrm{op}(\omega+\Omega,r,z)A^*_\mathrm{op}(\omega,r,z)d\omega \left[\frac{e^{i\delta z}-e^{\frac{-\alpha(\Omega) z}{2}}}{\frac{\alpha(\Omega)}{2}+i\delta}\right] \label{1d_thz}\label{1_d_thz}\\
=&&P^\mathrm{NL}(\Omega)\left[\frac{e^{i\delta z}-e^{\frac{-\alpha(\Omega) z}{2}}}{\frac{\alpha(\Omega)}{2}+i\delta}\right] \nonumber
\end{eqnarray}
where , 
\begin{align}
& \delta=[k(\Omega)+k(\omega)-k(\omega+\Omega)-\frac{2\pi}{\Lambda}]=[\frac{\Omega}{c}(n(\Omega)-n_\mathrm{g}(\omega))-\frac{2\pi}{\Lambda}]\approx \frac{ (n(\Omega_{0})-n_\mathrm{g}(\omega_0)) }{c \tau_{\text{FWHM}}}.
\end{align}

In order to study the distance $L_0$ and $L_\mathrm{eff}$, we calculate the second order derivative of $|A_\text{THz}|^2$ which is proportional to the efficiency $\eta$. By calculating ${d^2|A_\mathrm{THz}(\Omega,z)|^2}/{d^2z}=0 $, one can get
\begin{eqnarray}
  e^{-\frac{\alpha z}{2}}=\left[\frac{1}{2}\cos(\delta z)+2\frac{\delta}{\alpha} \sin(\delta z)-2\frac{\delta^2}{\alpha^2}\cos(\delta z) \right] \\
\begin{cases}
\text{for short pump pulse},~  \frac{\delta}{\alpha} \gg 1, 2\frac{\delta}{\alpha} \sin(\delta z)=2\frac{ \delta^2}{\alpha^2}\cos(\delta z) \\
\Rightarrow L_0={\tan^{-1}(\frac{\delta}{\alpha}})/{\delta}, \\
\ \nonumber \\
\text{for long pump pulse,} ~ \frac{\delta}{\alpha} \ll 1 ,~  e^{-\frac{\alpha z}{2}}=\frac{1}{2}\cos(\delta z)\approx\frac{1}{2}\\ \Rightarrow L_0={2\ln{(2)}}/{\alpha},L_\text{eff}=\frac{2}{\alpha}\ln{(\frac{2}{1-\sqrt{1-e^{-1}}})}
\end{cases} \nonumber
\end{eqnarray}
\end{document}